\documentclass[acmsmall,screen,nonacm]{acmart}
\ifdefined\pdfmapfile
  \pdfmapfile{+zi4.map}
\fi
\usepackage{listings}
\usepackage{tabularx}
\usepackage{booktabs}
\usepackage{multicol, multirow}
\usepackage{amsmath,amsfonts}
\usepackage{algorithmic}
\usepackage[ruled,vlined,linesnumbered]{algorithm2e}
\SetAlgoSkip{2pt}
\SetAlCapSkip{2pt}
\SetInd{0.5em}{0.5em}
\SetAlgoInsideSkip{smallskip}
\usepackage{graphicx}
\usepackage{textcomp}
\usepackage{xcolor}
\usepackage{colortbl}
\usepackage[frozencache]{minted}
\usepackage{tikz}
\usepackage{pgfplots}
\pgfplotsset{compat=1.18}
\usetikzlibrary{arrows.meta, positioning, shapes.geometric, decorations.pathreplacing, patterns, calc}
\usepackage{subcaption}
\usepackage{placeins}
\def\BibTeX{{\rm B\kern-.05em{\sc i\kern-.025em b}\kern-.08em
    T\kern-.1667em\lower.7ex\hbox{E}\kern-.125emX}}
\begin{document}

\title{Interrupt-Aware Static Detection of Deadlocks in Rust Operating Systems}

\author{Yuzhi Li}
\affiliation{%
  \institution{Fudan University}
  \city{Shanghai}
  \country{China}
}
\email{yuzhili24@m.fudan.edu.cn}

\author{Hongliang Tian}
\affiliation{%
  \institution{Ant Group}
  \city{Beijing}
  \country{China}
}
\email{tate.thl@antgroup.com}

\author{Chengjun Chen}
\affiliation{%
  \institution{Ant Group}
  \city{Beijing}
  \country{China}
}
\email{chenchengjun.ccj@antgroup.com}

\author{Jianfeng Jiang}
\affiliation{%
  \institution{Ant Group}
  \city{Beijing}
  \country{China}
}
\email{jiangjianfeng.jjf@antgroup.com}

\author{Hui Xu}
\affiliation{%
  \institution{Fudan University}
  \city{Shanghai}
  \country{China}
}
\email{xuh@fudan.edu.cn}

\begin{abstract}
Rust is increasingly adopted in operating systems, but Rust's safety guarantees do not eliminate deadlocks, especially those induced by interrupts. The core challenge is to recover interrupt-sensitive lock dependencies from Rust's guard-based synchronization across function boundaries. Applying this analysis across Rust OS projects also requires specifying project-specific synchronization semantics. We present an approach combining interrupt-aware static analysis with diagnostic-guided agent auditing for detecting deadlocks in Rust operating systems. Our approach uses lightweight source-level annotations to recover project-specific lock and interrupt semantics, performs a flow-sensitive MIR-based analysis to track lock-holding and interrupt states, and reports potential deadlocks with supporting evidence. An agent then searches for relevant source code in the repository and returns a confidence assessment with reasons for each supplied report. We evaluate our tool on four real Rust operating systems, identifying 26 true cases corresponding to nine maintainer-confirmed defects. Diagnostic-guided auditing achieves 86.2\% precision and 96.2\% recall based on static analysis results, and offers advantages over the baselines in both cost and effectiveness.
\end{abstract}



\maketitle

\section{Introduction}
\label{sec:introduction}

Rust combines low-level control with strong memory-safety guarantees, motivating adoption in systems software~\cite{balasubramanian2017system,fulton2021benefits}, including Linux, embedded systems, and research operating systems~\cite{li2024empirical,sharma2024rust,narayanan2020redleaf,boos2020theseus,peng2024framekernel,peng2025asterinas,tian2025asterinas}. Its ownership model rules out many memory errors and data races, but does not ensure that concurrent kernel operations make progress. Deadlocks can still stall critical kernel paths and I/O, compromising availability even when accesses are memory-safe. Empirical studies confirm that concurrency bugs remain a practical concern despite Rust's safety guarantees~\cite{qin2020rust,qin2024understanding}.

Kernel deadlock detection must account for lock interactions across call chains and execution contexts, including interrupt preemption. Acquisitions and releases span subsystem boundaries, while interrupt handlers introduce execution paths that are not explicit in the normal call sequence~\cite{li2023embedded}. If an interrupt handler preempts a lock holder on the same CPU and attempts to acquire the same nonrecursive lock, the handler waits while the suspended holder cannot resume to release it. Such deadlocks can occur without a lock-order inversion between ordinary threads. Their dependence on narrow interrupt timing windows makes them easy to overlook in manual review and difficult to reproduce through testing, motivating runtime validators and dedicated static kernel analyses~\cite{corbet2006lockdep,bai2022dlos,ye2024archerfish}.

Prior work provides two complementary foundations. Static kernel analyses recover lock dependencies, and interrupt-aware analyses model additional dependencies induced by interrupt preemption~\cite{bai2022dlos,ye2024archerfish}. Rust-specific analyses reason about lock misuse through ownership and guard lifetimes~\cite{ning2020stuckmenot,zhang2024rustdeadlock,qin2024understanding}. However, their combination remains a gap for Rust operating systems. Existing kernel analyses largely target C code with explicit lock/unlock operations, whereas Rust synchronization uses RAII (Resource Acquisition Is Initialization) guards whose ownership transfers and destruction determine resource lifetimes~\cite{jung2021safe}. An analysis must connect each guard to the concrete lock it protects and interpret its synchronization effects across function boundaries. Existing Rust deadlock analyses do not model kernel interrupt contexts or guard-induced IRQ changes: ignoring these effects can miss deadlocks, while treating all execution as interruptible introduces infeasible dependencies. Detecting deadlocks in this setting therefore requires jointly recovering which lock instances may remain held and where interrupts may be enabled, including when a callee changes IRQ state while its caller still holds a lock.

We present a tool combining interrupt-aware static analysis with diagnostic-guided agent auditing for Rust operating systems. To accommodate different synchronization primitives across projects, lightweight source-level annotations supply their semantics; a flow-sensitive analysis of MIR (Mid-level Intermediate Representation) recovers guard-managed lock and interrupt states, constructs an interrupt-aware lock dependency graph (LDG), and extracts candidate cycles with supporting evidence. Guided by these reports, an agent inspects repository source to assess whether the dependencies can arise together in an actual execution, then returns confidence labels and reasons for human review.

We implement this workflow in a prototype and evaluate it on four real Rust OS projects, identifying 26 true cases corresponding to nine maintainer-confirmed defects, with diagnostic-guided auditing assigning high confidence to 25 of those cases.

In summary, this paper makes the following contributions:

\begin{itemize}
    \item We present a guard-aware static analysis approach based on MIR that constructs an interrupt-aware lock dependency graph for detecting interrupt-induced deadlocks in Rust OS projects.
    \item We introduce an annotation-based mechanism that binds project-specific lock and interrupt primitives to the analysis. The annotations provide semantic metadata without hard-coding project-specific synchronization primitives into the analyzer.
    \item We implement a prototype combining static analysis with diagnostic-guided agent auditing and evaluate it on four real Rust OS projects, demonstrating real defect detection capability.
\end{itemize}

\section{Background}
\label{sec:background}

\subsection{Deadlock Patterns and Detecting Method}

The classical \emph{ABBA} deadlock occurs when one execution flow holds $A$ and waits for $B$, while another holds $B$ and waits for $A$~\cite{lu2008learning}. Static analyses recover locking dependencies and search for cycles in a lock dependency graph (LDG)~\cite{engler2003racerx,fu2018survey}. In our formulation, nodes represent lock instances; for ordinary nested locking, an edge $X \rightarrow Y$ indicates that $X$ may be acquired while $Y$ remains held. A cycle indicates a potential deadlock, rather than establishing a feasible execution.

In operating systems, an interrupt may preempt a lock holder and transfer control to an interrupt service routine (ISR) on the same CPU. If the ISR spins on the same nonrecursive lock, it cannot return, while the interrupted execution cannot resume to release the lock until the ISR returns. This produces an \emph{AA} self-deadlock without a classical lock-order inversion~\cite{linuxlockdoc,corbet2006lockdep}. Thread-centric analyses that treat these contexts as independently progressing execution flows do not capture the suspended owner's dependency on ISR completion. Each context may acquire $A$ only once, so neither contains an ordinary nested acquisition of $A$ from which to derive an $A \rightarrow A$ edge. Without modeling interrupt preemption, their ordinary nested-locking edges omit this self-dependency. Detecting this pattern therefore requires explicit interrupt modeling.

\subsection{Limitations of Existing Static Approaches }

Kernel deadlock analyses for C recover lock lifetimes from explicit \texttt{lock}/\texttt{unlock} APIs, with recent work also modeling interrupt-induced dependencies~\cite{bai2022dlos,ye2024archerfish}. This reliance on explicit operations does not recover Rust's guard-managed synchronization effects. Conversely, Rust deadlock analyses recover dependencies through ownership and guard lifetimes~\cite{ning2020stuckmenot,zhang2024rustdeadlock}, but do not model kernel interrupt preemption. Rust OS analysis requires combining these capabilities: a guard may carry both lock and IRQ effects, which must be interpreted together to determine whether a held lock can be re-entered by an ISR.

\subsection{Key Concepts}

A \emph{lock instance} is a concrete synchronization object. Acquiring it typically returns a \emph{lock guard instance}, whose destruction releases the lock. The holding interval ends at the guard's drop, which may be compiler-inserted, rather than necessarily at its last explicit use. Moving a guard transfers ownership without releasing its resource; its eventual drop may occur in another function.

We use \emph{interrupt state} to denote whether the current execution may be preempted by local hardware interrupts (IRQs). RAII-style \emph{IRQ guards} keep IRQs disabled while held and restore or update IRQ state according to their semantics when dropped. \emph{ISR-reachable} functions may execute directly or transitively along the call graph from an ISR entry.

\section{Motivating Examples}
\label{sec:motivating-examples}

\newsavebox{\exampletakeawaybox}
\newcommand{\exampletakeaway}[1]{%
  \par\addvspace{\smallskipamount}%
  \begingroup
  \setlength{\fboxsep}{6pt}%
  \sbox{\exampletakeawaybox}{%
    \parbox{\dimexpr\linewidth-2\fboxsep-2pt\relax}{#1}}%
  \noindent{\color{black!65}%
    \rule[-\dimexpr\dp\exampletakeawaybox+\fboxsep\relax]{2pt}%
      {\dimexpr\ht\exampletakeawaybox+\dp\exampletakeawaybox+2\fboxsep\relax}}%
  \colorbox{black!6}{\usebox{\exampletakeawaybox}}%
  \par
  \endgroup
  \addvspace{\smallskipamount}%
}

\begin{figure*}[t]
  \centering
  \includegraphics[width=\textwidth]{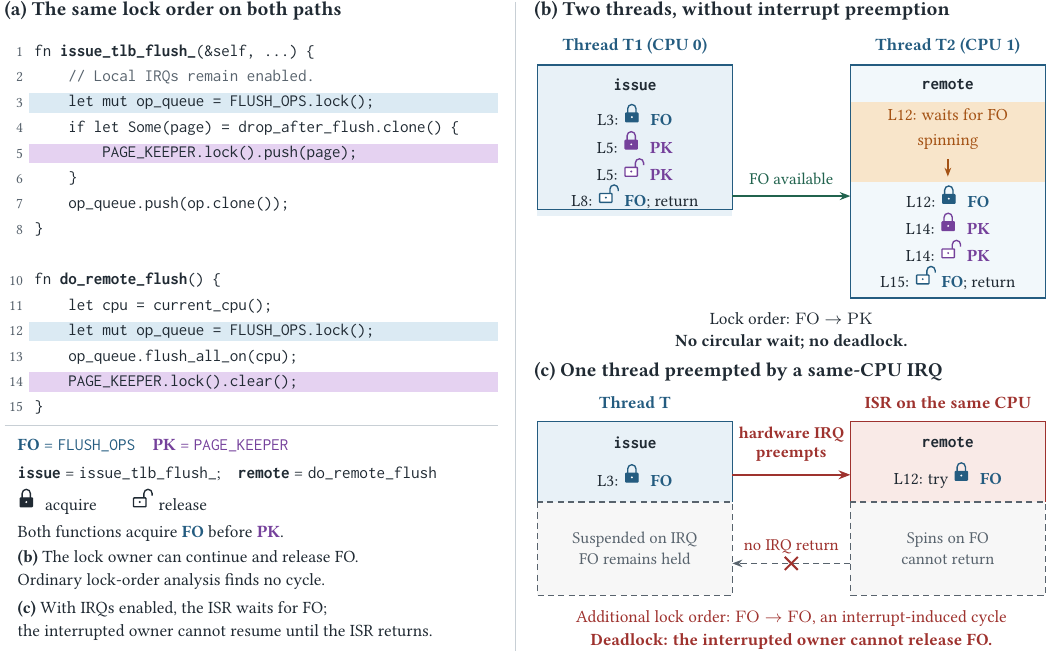}
  \input{figures/flush-irq-example/caption}
  \label{fig:example-code}
  \Description{Left: simplified Rust code for do_remote_flush and issue_tlb_flush_.
  Both acquire FLUSH_OPS and then PAGE_KEEPER without masking local interrupts.
  Upper right: two ordinary threads on different CPUs execute the functions. Thread T2 initially
  spins on FLUSH_OPS, but thread T1 continues and releases it; T2 then acquires
  both locks and finishes. Equal-width frames share horizontal event rows.
  Closed and open padlocks mark acquisition and release; blue is FLUSH_OPS and
  purple is PAGE_KEEPER. A waiting band spans the rows in which T1 works on
  PAGE_KEEPER, and an arrow connects T1's FLUSH_OPS release to T2's acquisition.
  The lock dependency graph is acyclic.
  Lower right: a hardware IRQ preempts the thread holding FLUSH_OPS and reaches
  do_remote_flush on the same CPU. The handler spins on the held lock. Grey
  dashed continuations and a crossed return arrow show that the handler cannot
  return and the thread cannot resume to release the lock. Interrupt modeling
  introduces a self-dependency on FLUSH_OPS.}
\end{figure*}

This section uses two motivating examples to illustrate why detecting interrupt-induced deadlocks in Rust operating systems is necessary and what makes static detection challenging.

\textbf{Traditional lock-order cycle detection misses interrupt-induced deadlocks.}
Figure~\ref{fig:example-code}(a) simplifies a TLB-flush example from a Rust OS. Both \texttt{issue\_tlb\_flush\_} (\emph{issue}) and \texttt{do\_remote\_flush} (\emph{remote}) acquire \texttt{FLUSH\_OPS} (FO) before \texttt{PAGE\_KEEPER} (PK). The locks do not mask local hardware interrupts, and \emph{remote} is reachable from an interrupt handler.

When these paths run as ordinary threads on different CPUs, waiting for FO does not prevent its owner from progressing and releasing it (Figure~\ref{fig:example-code}(b)). This conclusion changes when \emph{remote} runs in an ISR on the \emph{same CPU} as \emph{issue} (Figure~\ref{fig:example-code}(c)). An interrupt arriving while \emph{issue} holds FO suspends its owner; the handler then spins on FO and cannot return. The owner cannot resume to release FO until the handler returns, so neither can complete. An interrupt-induced self-dependency $\mathrm{FO}\rightarrow\mathrm{FO}$ captures this circular wait.

\exampletakeaway{A detector that considers only ordinary thread interleavings would miss this deadlock despite the consistent lock order. Detecting it requires modeling interrupt preemption and the resulting dependency on the suspended lock owner.}

\begin{figure*}[t]
  \centering
  \includegraphics[width=\textwidth]{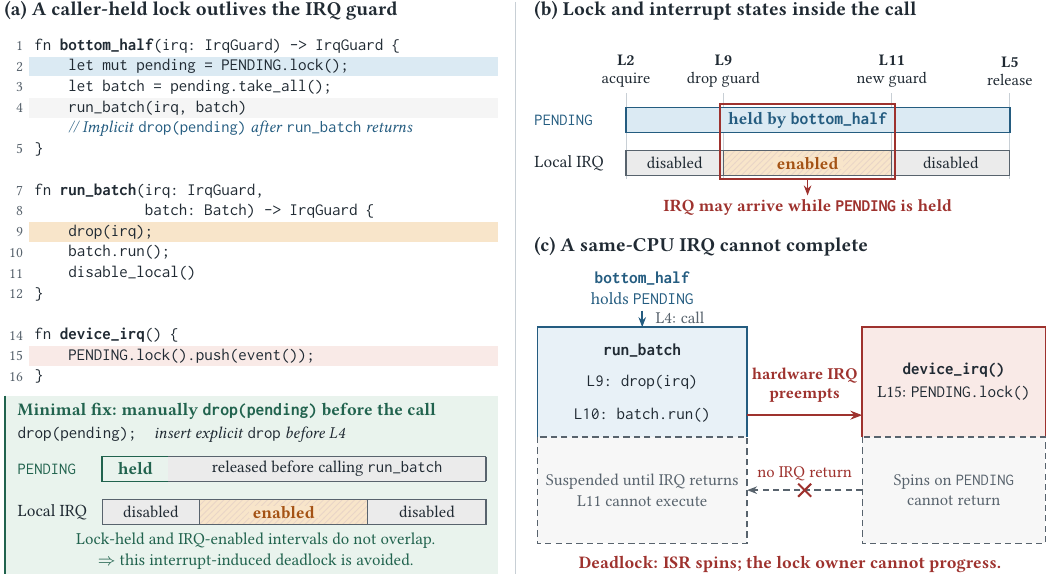}
  \input{figures/guard-irq-example/caption}
  \label{fig:guard-irq-example}
  \Description{Left: Rust code keeps a queue lock in bottom_half while run_batch
  consumes and replaces an interrupt guard. Below the code, a repair box combines
  the early drop with aligned queue and local-IRQ state bars. The queue is released
  before the disabled-enabled-disabled IRQ sequence reaches its enabled interval.
  Top right: the queue remains locked
  throughout an interval in which local interrupts are enabled, although the
  callee enters and returns with interrupts disabled. Bottom right: on one CPU,
  bottom_half calls run_batch, which drops the IRQ guard and executes batch.run.
  A hardware interrupt preempts that execution and enters device_irq. The handler
  spins on the queue lock; the interrupted context remains suspended with the lock
  held, unable to reach the new IRQ guard or implicit queue release. An early drop
  of the queue guard before the call removes the overlap.}
\end{figure*}

\textbf{Modeling interrupt and lock states requires tracking guard lifetimes across functions.}
Figure~\ref{fig:guard-irq-example}(a) illustrates a bottom-half processing protocol commonly used in OS kernels: local IRQs are disabled on entry and exit but enabled during batch processing. Here, \texttt{bottom\_half} extracts an owned batch from \texttt{PENDING} and passes it with the incoming IRQ guard to \texttt{run\_batch}, which drops the guard before processing and returns a replacement. We introduce an illustrative defect by retaining the queue lock across this call.

Assume that \texttt{PENDING} is a nonrecursive spinlock that does not mask IRQs, the incoming IRQ guard restores Enabled when dropped, and \texttt{batch.run()} neither changes IRQ state nor accesses the queue. Although \texttt{pending} is last used explicitly at L3, its implicit drop occurs at L5, after the call at L4 returns~\cite{rustReferenceDestructors}. Along the uninterrupted path, the lock-held interval L2--L5 overlaps the IRQ-enabled window between dropping \texttt{irq} at L9 and creating its replacement at L11 (Figure~\ref{fig:guard-irq-example}(b)). A device IRQ arriving on the same CPU can then attempt to acquire the same lock at L15 (Figure~\ref{fig:guard-irq-example}(c)). The handler spins on the suspended caller's lock, preventing the caller from resuming to release it.

Although \texttt{run\_batch} enters and returns with IRQs disabled, its internal transition must be interpreted together with the lock still held by its caller.

\exampletakeaway{In order to model interrupt and lock states, a detector must track guards across function call boundaries and recover their associated lock and interrupt states at each program point.}

The following section summarizes the resulting analysis challenges and our solutions.

\section{Challenges and Solutions}
\label{sec:challenges}

The motivating examples raise two challenges: recovering synchronization effects from RAII guards across calls, and obtaining the project-specific primitive semantics needed to apply this analysis across Rust OSes.

\subsection{Challenge 1: Detecting Interrupt-Induced Deadlocks under Guard-Based Synchronization}
\label{sec:challenge-guards}

Prior kernel analyses model interrupt-induced dependencies~\cite{ye2024archerfish}, while Rust analyses recover lock dependencies from ownership and guard lifetimes~\cite{zhang2024rustdeadlock}; neither jointly recovers guard-managed lock and IRQ effects across calls. The central challenge is to connect guard lifetimes to their synchronization effects in the calling context. Locating lifetime endpoints alone does not reveal those effects. Moving a guard into a callee transfers ownership without releasing its resource; destroying it may release a lock, restore a saved IRQ state, or do both. These effects must be associated with the particular guard value and propagated across function boundaries, so that a callee's IRQ transition is interpreted together with locks still held by its caller.

\noindent\textbf{Solution.}
Our approach jointly recovers lock and interrupt states from guard lifecycles through a flow-sensitive, inter-procedural analysis of MIR. It associates guard values with the locks they protect and their IRQ effects, propagates these associations across calls, and interprets guard acquisition and destruction according to the primitive's semantics. This recovers where locks may remain held and where IRQs may be enabled. These states support an interrupt-aware lock dependency graph that combines ordinary locking dependencies with those induced by ISR acquisitions (Sections~\ref{sec:approach-analysis} and~\ref{sec:approach-ldg}).

\subsection{Challenge 2: Adapting to Project-Specific Synchronization Semantics}
\label{sec:challenge-semantics}

Applying this analysis requires knowing each target's synchronization primitives and their effects. Linux-focused detectors can build on a known kernel primitive vocabulary~\cite{bai2022dlos,ye2024archerfish}. Across Rust OSes, however, locks may come from third-party crates, local wrappers, or custom implementations, and IRQ control may use explicit operations or guards. Obtaining these project-specific semantics through configuration avoids hard-coding each target's APIs into the detector.

\begin{figure}[ht]
  \input{figures/lock-irq-annotations/content}
  \input{figures/lock-irq-annotations/caption}
  \label{fig:lock-irq-annotations}
  \Description{Two side-by-side Rust listings show the shared types and Drop
  implementation on the left and the Lock implementation on the right.
  The Guard type contains an optional saved interrupt state. The shared
  destructor unlocks first and then restores that state if present.
  The lock method returns a guard with
  no saved state and has GuardIrqDisabled false. The lock_irqsave method saves
  and disables interrupts before acquisition, returns a guard with that saved
  state, and has GuardIrqDisabled true. Both attributes identify self as the
  acquired lock. The same guard type therefore has different IRQ effects
  depending on the acquisition operation.}
\end{figure}

Simply labeling API names as lock or interrupt operations is insufficient. Figure~\ref{fig:lock-irq-annotations} shows that the same guard type can be returned by acquisitions that either leave IRQs unchanged or mask them for the guard's lifetime. Configuration must therefore connect the acquisition operation to its lock instance, returned guard, and IRQ behavior, so that the required states can be recovered for the analysis.

\noindent\textbf{Solution.}
We use semantic annotations expressed as source-level attributes (Section~\ref{sec:approach-infocollection}). They bind acquisition operations to their lock operands, returned guards, and IRQ effects, providing the analysis with synchronization contracts. Attaching IRQ semantics to acquisitions distinguishes effects even when guard types coincide. These contracts keep project-specific configuration separate from detector code: projects can adapt covered primitives without modifying the detector. Keeping annotations with the source also lets projects version and review them alongside code changes and integrate checks into CI.

\section{Approach}
\label{sec:approach}
\subsection{Workflow Overview}
\label{sec:approach-overview}

\begin{figure}[t]
  \centering
  \includegraphics[width=\linewidth]{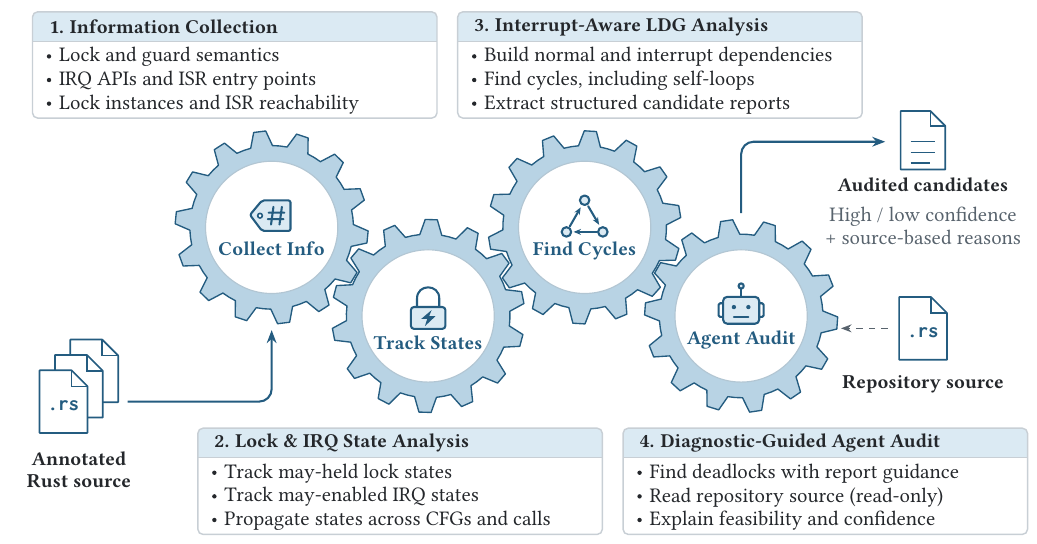}
  \input{figures/workflow/caption}
  \label{fig:workflow}
  \Description{Four blue gears form an interlocking chain from left to right.
  Their centers pair an attribute-tag icon with Collect Info, a lock and
  lightning icon with Track States, a directed-cycle icon with Find Cycles,
  and a robot icon with Agent Audit. Numbered surrounding panels describe information collection,
  lock and interrupt state analysis, interrupt-aware lock dependency graph
  analysis, and diagnostic-guided agent auditing. A stack of three source files,
  with .rs on the front file, enters stage one from below. A matching rounded
  right-angle path leaves stage four from above for the audited candidates.
  Repository source is a separate read-only input to stage four, shown with
  a file icon matching the size of each source-stack page.
  The final output consists of audited candidates with high or low confidence
  and source-based reasons.}
\end{figure}

The workflow consists of four stages (Figure~\ref{fig:workflow}). Stage~1 collects lock- and interrupt-related metadata from annotated source code. Stage~2 performs a flow-sensitive state analysis to compute which locks may be held and whether interrupts may be enabled at each program point. Stage~3 constructs an interrupt-aware lock dependency graph (LDG), searches for cycles, and extracts their supporting analysis facts into structured deadlock candidate reports. Stage~4 audits the candidates against repository source code, returning a high/low confidence label with reasons for each report. Together, these stages produce individually assessed candidates whose static evidence and source-based explanations remain available for review.

\subsection{Information Collection}
\label{sec:approach-infocollection}

The first stage collects project-specific concurrency-related semantics. Our tool obtains this information from source-level attributes, which serve as lightweight configuration embedded directly in the target repository. These attributes are attached manually by tool users; Section~\ref{sec:eval-rq3} evaluates annotation effort across Rust OS projects.

\begin{table}[t]
\caption{Custom attribute tags for lock and interrupt semantics}
\label{tab:attr-tags}
\centering
\small
\renewcommand{\arraystretch}{1.15}
\setlength{\tabcolsep}{4pt}
\setlength{\extrarowheight}{2pt}
\begin{tabularx}{\linewidth}{>{\raggedright\arraybackslash}p{0.44\linewidth}>{\raggedright\arraybackslash}X}
\noalign{\hrule height 1pt}
\textbf{Attribute tag} & \textbf{Semantics} \\[2pt]
\noalign{\hrule height 1pt}
\texttt{\#[\textbf{LockType}\detokenize{(Name = "xxx")]}} &
Marks a lock \texttt{struct}; the user-defined \texttt{Name} links related metadata. \\[2pt]
\hline
\texttt{\#[\textbf{LockGuardType}\detokenize{(Name = "xxx")]}} &
Marks a guard \texttt{struct} corresponding to a lock type. \\[2pt]
\hline
\texttt{\#[\textbf{LockOp}\detokenize{(LockArg = N,}}\newline
\hspace*{1em}\texttt{\detokenize{GuardIrqDisabled = true|false)]}} &
Marks a guard-producing lock acquisition. \texttt{LockArg} identifies the target lock argument; \texttt{GuardIrqDisabled} records whether the guard keeps local IRQs disabled while held. \\[2pt]
\hline
\texttt{\#[\textbf{IntrApi}\detokenize{(Type = "Enable" | "Disable")]}} &
Marks an interrupt-control function; \texttt{Type} selects enable or disable. \\[2pt]
\hline
\texttt{\#[\textbf{IsrEntry}\detokenize{(Name = "xxx")]}} &
Marks a function as an interrupt entry point. \\[2pt]
\noalign{\hrule height 1pt}
\end{tabularx}
\end{table}

Table~\ref{tab:attr-tags} summarizes the attributes used by the current prototype. From these attributes, the tool constructs metadata for state analysis and dependency construction: lock instances, mappings from guard-producing acquisitions and returned guards to those instances, acquisition-specific held-IRQ semantics, and explicit interrupt-control APIs. The tool traverses the call graph from annotated ISR entries to compute ISR-reachable functions. In the motivating example, this identifies \verb|FO| and \verb|PK| as lock instances and \verb|remote| as ISR-reachable.

\subsection{Lock and Interrupt State Analysis}
\label{sec:approach-analysis}

The second stage performs a forward, flow-sensitive analysis over MIR. At each program point it tracks $\mathit{ProgramState} = \mathit{IrqState} \times \mathit{LockSet}$, where \textit{IrqState} records whether interrupts may be enabled, and \textit{LockSet} maps each recovered lock instance to whether it may be held. Figure~\ref{fig:state-transfer-example}(a) summarizes the domains and transfer rules. $Enabled$ and $Held$ conservatively express possibilities; $Disabled$ and $Released$ indicate that the corresponding possibility is absent.

A tagged lock acquisition marks the corresponding lock instance as held and associates the returned guard with that instance. IRQ-disabling acquisitions also update \textit{IrqState}. Ownership transfers preserve these associations: moving a guard into a callee does not release its lock or remove its IRQ constraint. Guard destruction releases any associated lock and updates or restores \textit{IrqState} according to the primitive's semantics. Explicit IRQ-control operations directly update \textit{IrqState}.

States propagate along MIR control-flow graphs (CFGs). At CFG joins, a conservative merge retains $Held$ for a lock if either incoming state may hold it, and $Enabled$ if either allows interrupts. Inter-procedural propagation uses bounded context sensitivity to interpret callee effects in the caller's synchronization context. On return, it applies callee exit effects, including releases and IRQ changes, and conservatively combines possible return states. The resulting may-held and may-enabled states identify potential overlaps between lock-holding intervals and interruptible execution.

Figure~\ref{fig:state-transfer-example}(b) shows this analysis procedure on the second motivating example(Figure~\ref{fig:guard-irq-example}). Assuming IRQs are enabled on entry to \texttt{handler}, dropping the incoming IRQ guard in \texttt{run\_batch} enables IRQs at line~9 while \texttt{PENDING} remains held; line~11 disables IRQs and returns a replacement guard. Although the call at line~4 starts and returns with IRQs disabled (R7), analyzing the callee's internal states in the caller's locking context exposes this overlap; R8 conservatively merges \texttt{handler}'s branches at line~20. These states supply the holding and interrupt contexts used in LDG construction.

\begin{figure*}[!t]
\centering
\begingroup
\captionsetup[subfigure]{skip=5pt}
\begin{subfigure}{\linewidth}
\small
\renewcommand{\arraystretch}{1.15}
\setlength{\tabcolsep}{4pt}
\begin{tabularx}{\linewidth}{l>{\raggedright\arraybackslash}X}
\noalign{\hrule height 1pt}
\multicolumn{2}{l}{\textbf{State domains}} \\
\hline
\multicolumn{2}{l}{$\mathit{ProgramState} = \mathit{IrqState} \times \mathit{LockSet}$} \\
\multicolumn{2}{l}{$\mathit{IrqState} = \{Enabled,\ Disabled\}$} \\
\multicolumn{2}{l}{$\mathit{LockSet} : \mathit{LockInstance} \rightarrow \{Held,\ Released\}$} \\
\noalign{\hrule height 1pt}
\textbf{Transfer rule} & \textbf{State update} \\
\noalign{\hrule height 1pt}
\textbf{R1}\quad Lock acquisition & \texttt{g = l.lock()} $\Rightarrow$ $\mathit{LockSet}[l] \leftarrow Held$ \\
\hline
\textbf{R2}\quad Acquisition + IRQ update & R1 with \texttt{GuardIrqDisabled = true} $\Rightarrow$ also set $\mathit{IrqState} \leftarrow Disabled$ \\
\hline
\textbf{R3}\quad Lock release & \texttt{drop(g)} $\Rightarrow$ $\mathit{LockSet}[l] \leftarrow Released$ \\
\hline
\textbf{R4}\quad Release + IRQ update & R3 for an IRQ-disabling guard; update $\mathit{IrqState}$ per guard semantics \\
\hline
\textbf{R5}\quad Explicit IRQ enable & \texttt{enable\_irq()} $\Rightarrow$ $\mathit{IrqState} \leftarrow Enabled$ \\
\hline
\textbf{R6}\quad Explicit IRQ disable & \texttt{disable\_irq()} $\Rightarrow$ $\mathit{IrqState} \leftarrow Disabled$ \\
\hline
\textbf{R7}\quad Function call return & Apply callee exit effects to caller; conservatively merge returns \\
\hline
\textbf{R8}\quad CFG join & \texttt{merge(s1, s2)} $\Rightarrow$ conservative merge of both state components \\
\noalign{\hrule height 1pt}
\end{tabularx}
\caption{Abstract-state domains and transfer rules.}
\label{fig:state-transfer-rules}
\end{subfigure}
\par\vspace{14pt}
\begin{subfigure}{\linewidth}
\centering
\renewcommand{\ttdefault}{zi4}
\definecolor{stateCodeGreen}{RGB}{0,115,15}
\definecolor{stateCodeBlue}{RGB}{0,52,194}
\definecolor{stateCodeComment}{RGB}{48,119,122}
\definecolor{stateInkBlue}{HTML}{245D7C}
\definecolor{statePaleBlue}{HTML}{EAF2F7}
\definecolor{stateIrqOrange}{HTML}{9A4D10}
\definecolor{statePaleOrange}{HTML}{FAE7CD}
\definecolor{stateMuted}{HTML}{555E64}
\resizebox{\linewidth}{!}{%
\begin{tikzpicture}[x=1cm,y=-1cm,
  every node/.style={inner sep=0pt,outer sep=0pt},
  code/.style={anchor=base west,font=\ttfamily\fontsize{8.3}{10}\selectfont},
  num/.style={anchor=base east,font=\rmfamily\fontsize{6.5}{8}\selectfont,text=stateMuted},
  note/.style={anchor=base west,font=\rmfamily\fontsize{8.3}{10}\selectfont},
  head/.style={anchor=base west,font=\rmfamily\bfseries\fontsize{10.2}{12}\selectfont}]
\node[head] at (.46,.35) {Code example};
\node[head] at (9.10,.35) {Rule / state effect};
\draw[line width=.5pt] (.46,.65) -- (17.55,.65);
\node[num] at (.27,1.020) {1};
\node[code] at (.46,1.020) {\textcolor{stateCodeGreen}{\textbf{fn}}\ \textcolor{stateCodeBlue}{bottom\_half}(irq:\ IrqGuard)\ -\textgreater{}\ IrqGuard\ \{};
\node[num] at (.27,1.335) {2};
\node[code] at (.46,1.335) {\ \ \ \ \textcolor{stateCodeGreen}{\textbf{let}}\ \textcolor{stateCodeGreen}{\textbf{mut}}\ pending\ =\ PENDING.lock();};
\node[note] at (9.10,1.335) {\textbf{R1}\quad \texttt{PENDING} $\leftarrow Held$};
\node[num] at (.27,1.650) {3};
\node[code] at (.46,1.650) {\ \ \ \ \textcolor{stateCodeGreen}{\textbf{let}}\ batch\ =\ pending.take\_all();};
\fill[statePaleBlue] (.42,1.735) rectangle (17.55,2.050);
\node[num] at (.27,1.965) {4};
\node[code] at (.46,1.965) {\ \ \ \ run\_batch(irq,\ batch)};
\node[note,text=stateInkBlue] at (9.10,1.965) {\textbf{R7}\quad IRQ at call / return: \textbf{Disabled / Disabled}};
\node[num] at (.27,2.280) {5};
\node[code] at (.46,2.280) {\}};
\node[note] at (9.10,2.280) {\textbf{R3}\quad Release \texttt{PENDING} after L4 returns};
\node[num] at (.27,2.910) {7};
\node[code] at (.46,2.910) {\textcolor{stateCodeGreen}{\textbf{fn}}\ \textcolor{stateCodeBlue}{run\_batch}(irq:\ IrqGuard,};
\node[num] at (.27,3.225) {8};
\node[code] at (.46,3.225) {\ \ \ \ \ \ \ \ \ \ \ \ \ batch:\ Batch)\ -\textgreater{}\ IrqGuard\ \{};
\fill[statePaleOrange] (.42,3.310) rectangle (17.55,3.625);
\node[num] at (.27,3.540) {9};
\node[code] at (.46,3.540) {\ \ \ \ drop(irq);};
\node[note,text=stateIrqOrange] at (9.10,3.540) {\textbf{R4}\quad Restore saved IRQ: \textbf{Enabled}};
\fill[statePaleOrange] (.42,3.625) rectangle (17.55,3.940);
\node[num] at (.27,3.855) {10};
\node[code] at (.46,3.855) {\ \ \ \ batch.run();};
\node[note,text=stateIrqOrange] at (9.10,3.855) {\texttt{PENDING} remains \textbf{Held} while IRQ is \textbf{Enabled}};
\node[num] at (.27,4.170) {11};
\node[code] at (.46,4.170) {\ \ \ \ disable\_local()};
\node[note] at (9.10,4.170) {\textbf{R6}\quad IRQ $\leftarrow Disabled$; return a new guard};
\node[num] at (.27,4.485) {12};
\node[code] at (.46,4.485) {\}};
\node[num] at (.27,5.115) {14};
\node[code] at (.46,5.115) {\textcolor{stateCodeGreen}{\textbf{fn}}\ \textcolor{stateCodeBlue}{handler}(is\_timer:\ \textcolor{stateCodeGreen}{\textbf{bool}})\ \{};
\node[num] at (.27,5.430) {15};
\node[code] at (.46,5.430) {\ \ \ \ \textcolor{stateCodeGreen}{\textbf{let}}\ \textcolor{stateCodeGreen}{\textbf{mut}}\ irq\ =\ disable\_local();};
\node[note] at (9.10,5.430) {\textbf{R6}\quad IRQ $\leftarrow Disabled$; create IRQ guard};
\node[num] at (.27,5.745) {16};
\node[code] at (.46,5.745) {\ \ \ \ \textcolor{stateCodeGreen}{\textbf{if}}\ is\_timer\ \{};
\node[num] at (.27,6.060) {17};
\node[code] at (.46,6.060) {\ \ \ \ \ \ \ \ irq\ =\ bottom\_half(irq);};
\node[note] at (9.10,6.060) {\textbf{R7}\quad Function call return};
\node[num] at (.27,6.375) {18};
\node[code] at (.46,6.375) {\ \ \ \ \}\ \textcolor{stateCodeGreen}{\textbf{else}}\ \{};
\node[num] at (.27,6.690) {19};
\node[code] at (.46,6.690) {\ \ \ \ \ \ \ \ handle\_device();};
\node[num] at (.27,7.005) {20};
\node[code] at (.46,7.005) {\ \ \ \ \}};
\node[note] at (9.10,7.005) {\textbf{R8}\quad CFG join};
\node[num] at (.27,7.320) {21};
\node[code] at (.46,7.320) {\ \ \ \ drop(irq);};
\node[num] at (.27,7.635) {22};
\node[code] at (.46,7.635) {\}};
\draw[line width=.5pt] (.46,7.925) -- (17.55,7.925);

\end{tikzpicture}%
}\par
\caption{Code example with corresponding rule applications. R2 is not illustrated.}
\label{fig:state-transfer-code}
\end{subfigure}
\endgroup
\input{figures/approach-state-transfer/caption}
\Description{Two vertically stacked subfigures: (a) a table presents the state domains and eight transfer rules; (b) Rust functions bottom_half, run_batch, and handler illustrate IRQ-guard transfer across calls and an if-else CFG join. Rules are aligned to source lines: R1 at line 2, R7 at lines 4 and 17, R3 at line 5, R4 at line 9, R6 at lines 11 and 15, and R8 at line 20. Blue shading marks IRQs disabled at both call and return on line 4; orange shading marks the internal IRQ-enabled interval on lines 9 and 10 while PENDING remains held. R2 is not illustrated.}
\label{fig:state-transfer-example}
\end{figure*}

\subsection{Interrupt-Aware LDG Construction and Analysis}
\label{sec:approach-ldg}

The third stage builds the interrupt-aware LDG from the preceding analysis, identifies candidate cycles, and extracts their supporting evidence. Nodes represent lock instances, and edges point from a new acquisition to a lock that may already be held.

\emph{Normal edges} represent nested locking. If the state analysis shows that $Y$ may still be held when $X$ is acquired, the tool adds $X \rightarrow Y$.

\emph{Interrupt edges} combine ISR reachability with lock and interrupt states. If an ISR-reachable function may acquire $X$, and the analysis finds a point where $Y$ may be held with interrupts enabled, the tool adds $X \rightarrow Y$. This captures the possibility of the ISR trying to acquire $X$ while the interrupted execution still holds $Y$.

Edge collection focuses on lock acquisitions and locations where the abstract state may change. Each edge retains the relevant acquisition and holding locations and its normal or interrupt context, so that the subsequent report can identify the source evidence underlying the dependency.

\begin{figure}[t]
\centering
\includegraphics[width=\linewidth]{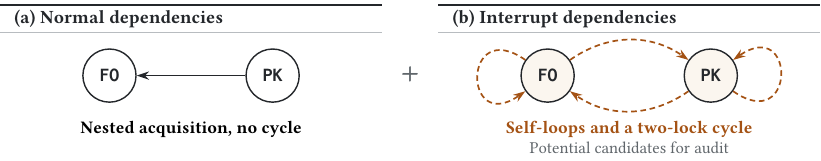}
\input{figures/ldg-example/caption}
\Description{Two panels show edge sets on the same two lock vertices, FO and PK,
with a plus sign indicating their union. The left panel has one solid normal
edge from PK to FO and no cycle. The right panel has four dashed interrupt
edges: FO to PK, PK to FO, and one self-loop on each vertex. The self-loops and
two-lock cycle are potential candidates for audit, not confirmed deadlocks.}
\label{fig:ldg-example}
\end{figure}

Figure~\ref{fig:ldg-example} decomposes the LDG for the first motivating example(Figure~\ref{fig:example-code}) into normal and interrupt dependencies. The ISR-reachable function \verb|remote| acquires both \verb|FO| and \verb|PK|, while \verb|issue| may hold either lock with interrupts enabled. The tool therefore adds a dependency for each combination, including self-loops. Unlike ordinary lock-order analysis, the interrupt edges also expose the possible dependency on a suspended lock owner.

After constructing the LDG, the tool searches for cycles, including self-loops, and extracts the supporting evidence for each candidate. It produces a structured candidate report containing an identifier, the involved lock instances, dependency edges and their types, the relevant acquisition and holding locations, and the lock and IRQ context supporting those dependencies. These facts explain why the static analysis emitted the candidate, but do not guarantee a feasible deadlock or constitute a complete execution trace. The report's source locations provide entry points for agent inspection, while its dependencies and states identify conditions that need to hold together.

\subsection{Diagnostic-Guided Agent Auditing}
\label{sec:approach-audit}

The fourth stage turns the structured candidates from LDG analysis into a project-level agent audit. The tool supplies all static candidate reports (\emph{diagnostics}), including their identifiers, to an agent run session. The prompt asks the agent to find all possible deadlocks in the target repository and assess their feasibility, returning a separate assessment for each report.

\begin{figure}[!t]
  \centering
  \includegraphics[width=\linewidth]{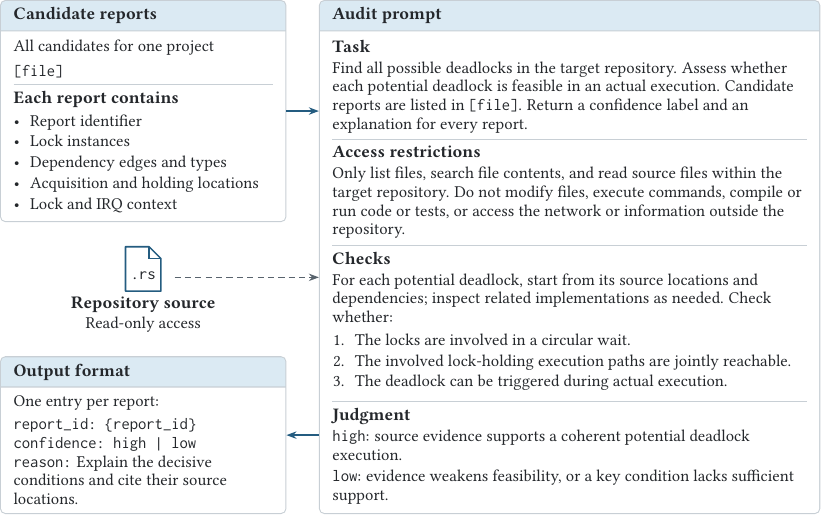}
  \input{figures/audit-prompt/caption}
  \Description{Two cards with blue header backgrounds on the left show the candidate-report file
  placeholder with a summary of each report's contents and the required output
  fields for each report: report identifier, confidence, and reason. Arrows connect them to the audit
  prompt on the right, which contains the task, repository-only access
  restrictions, three checks covering circular wait, joint reachability of
  lock-holding paths, and runtime triggering, and high and low judgment criteria.
  A separate Rust source-file icon connects to the prompt with a dashed arrow
  indicating read-only repository access.}
  \label{fig:audit-prompt}
\end{figure}

The agent can only list files, search file contents, and read source files within the target repository, without modification, execution, or network access. For each candidate, it returns \texttt{report\_id}, \texttt{confidence}, and \texttt{reason}, preserving a one-to-one correspondence between candidate reports and completed audit assessments. Starting from the reported source locations and dependencies, it can follow callers and callees, examine guard or synchronization implementations, and consult interrupt registration or initialization code.

The audit applies three checks: whether the locks are involved in a circular wait, whether the involved lock-holding execution paths are jointly reachable, and whether the deadlock can be triggered during actual execution. A \texttt{high} label means that the source evidence supports a coherent potential deadlock execution; a \texttt{low} label means that evidence weakens its feasibility or that a key condition lacks sufficient support. The explanation identifies the decisive conditions and cites their source locations. These source-based confidence labels support subsequent human review.

\FloatBarrier

\section{Evaluation}
\label{sec:evaluation}

This section evaluates how the complete workflow supports detection and review across Rust OS projects through experiments on detection effectiveness, baseline and stage contributions, resource costs, and annotation effort.

\subsection{Research Questions}
\label{sec:eval-rq}

We organize the evaluation around three questions:
\begin{itemize}
    \item \textbf{RQ1 (Detection effectiveness):} Can our tool identify real deadlock cases in Rust operating systems and provide useful confidence assessments?
    \item \textbf{RQ2 (Baseline comparison and stage contributions):} How does our tool compare with state-of-the-art static analyzer and agent-only auditing, in detection effectiveness and resource consumption? What does each stage of our tool contribute?
    \item \textbf{RQ3 (Annotation effort):} How much annotation effort is needed to configure each project, and how many lock and guard instances does the analyzer recover from this configuration?
\end{itemize}

\subsection{Experimental Setup}
\label{sec:eval-setup}

The experiments are run on an x86 machine with an Intel i7-9700T CPU and 32\,GB of RAM. We evaluate four popular open-source Rust OS projects on GitHub: \textsc{Asterinas}, \textsc{DragonOS}, \textsc{Motor-OS}, and \textsc{Moros}. Table~\ref{tab:eval-static-resources} reports their code sizes and static-analysis resource usage.

\emph{Agent configuration.}
Both agent workflows use GPT-5.6 Sol with high reasoning effort through the Codex harness, with isolated runs and external tools disabled. They receive the same prompt and repository source; the complete workflow additionally receives static deadlock candidates. Agents may list files, search their contents, and read repository source; they cannot access the network, modify files, execute commands, compile code, or run tests. No token cap or timeout is imposed; each session ends when the agent completes its response.

\emph{Validation and counting.}
We manually check the reported cases against source code. Cases marked as true positives correspond to defects confirmed by project maintainers, while those marked as false positives are infeasible deadlocks. The agent's \texttt{high}/\texttt{low} confidence labels are evaluated against the human judgments.

\emph{Cost measurements.}
We measure execution time, peak memory usage, and agent token consumption, estimating reference token costs using the rates in Figure~\ref{fig:eval-costs}.

\subsection{RQ1: Detection Effectiveness}
\label{sec:eval-rq1}

For each project, we supply the semantic annotations, run the static analysis, and pass the resulting reports to the auditing agent. We then examine the complete candidate set and its high-confidence subset using the case-level judgments defined above.

\begin{table}[!htbp]
\centering
\caption{Detection results of the full pipeline.}

\begingroup
\fontsize{9.0}{10.5}\selectfont
\renewcommand{\arraystretch}{1.05}
\setlength{\extrarowheight}{0pt}
\setlength{\aboverulesep}{1.5pt}
\setlength{\belowrulesep}{2pt}
\setlength{\tabcolsep}{3.2pt}
\noindent
\begin{minipage}{\linewidth}
\begin{tabular*}{\linewidth}{@{\extracolsep{\fill}}lrrrrr@{}}
\toprule
\multirow{2}{*}{Target} & \multicolumn{2}{c}{Static analysis result} & \multicolumn{2}{c}{Agent audit result} & \multirow{2}{*}{Distinct Bugs} \\
\cmidrule(lr){2-3}\cmidrule(lr){4-5}
 & Cases & True cases & High-confidence cases & True cases & \\
\midrule
\textsc{Asterinas} & 21 & \cellcolor{black!8}10 & 9 & \cellcolor{black!8}9 & 4 \\
\textsc{DragonOS} & 70 & \cellcolor{black!8}10 & 14 & \cellcolor{black!8}10 & 2 \\
\textsc{Motor-OS} & 2 & \cellcolor{black!8}2 & 2 & \cellcolor{black!8}2 & 1 \\
\textsc{Moros} & 39 & \cellcolor{black!8}4 & 4 & \cellcolor{black!8}4 & 2 \\
\midrule
\textbf{Total} & \textbf{132} & \cellcolor{black!8}\textbf{26} & \textbf{29} & \cellcolor{black!8}\textbf{25} & \textbf{9} \\
\bottomrule
\end{tabular*}
\par\vspace{3pt}
{\fontsize{8}{10}\selectfont\noindent \par}
\end{minipage}
\endgroup

\label{tab:eval-effectiveness}
\end{table}

\emph{Detection results.}
Table~\ref{tab:eval-effectiveness} shows that our tool produces 132 candidates across the four projects. Of these, 26 are true cases corresponding to nine distinct, maintainer-confirmed defects.

The agent assigns high confidence to 29 cases, including 25 of the 26 true cases. The effect of auditing on false positives and manual review is examined in RQ2.

\begin{table}[!htbp]
\centering
\caption{Project code size and static-analysis time and memory compared with \texttt{cargo check}.}

\begingroup
\fontsize{9.2}{10.5}\selectfont
\renewcommand{\arraystretch}{1.05}
\setlength{\extrarowheight}{0pt}
\setlength{\aboverulesep}{1.5pt}
\setlength{\belowrulesep}{2pt}
\setlength{\tabcolsep}{3.2pt}
\noindent
\begin{minipage}{\linewidth}
\begin{tabular*}{\linewidth}{@{\extracolsep{\fill}}lrrrrr@{}}
\toprule
 &  & \multicolumn{2}{c}{Execution time (s)} & \multicolumn{2}{c}{Peak memory (MB)} \\
\cmidrule(lr){3-4}\cmidrule(l){5-6}
Target & KLoC & Static analysis stage & \texttt{cargo check} & Static analysis stage & \texttt{cargo check} \\
\midrule
\textsc{Asterinas} & 96.7 & 110.2 & \cellcolor{black!8}35.3 & 911 & \multicolumn{1}{>{\columncolor{black!8}[\tabcolsep][0pt]}r@{}}{699} \\
\textsc{DragonOS} & 202.0 & 102.1 & \cellcolor{black!8}54.0 & 2,557 & \multicolumn{1}{>{\columncolor{black!8}[\tabcolsep][0pt]}r@{}}{1,520} \\
\textsc{Motor-OS} & 67.6 & 41.3 & \cellcolor{black!8}32.1 & 960 & \multicolumn{1}{>{\columncolor{black!8}[\tabcolsep][0pt]}r@{}}{968} \\
\textsc{Moros} & 21.1 & 19.5 & \cellcolor{black!8}16.8 & 961 & \multicolumn{1}{>{\columncolor{black!8}[\tabcolsep][0pt]}r@{}}{964} \\
\bottomrule
\end{tabular*}
\par\vspace{3pt}
{\fontsize{8}{10}\selectfont\noindent KLoC denotes thousands of lines of code.\par}
\end{minipage}
\endgroup

\label{tab:eval-static-resources}
\end{table}

\emph{Static-analysis cost.}
Static analysis takes 1.16--3.12 times the corresponding \texttt{cargo check} time (Table~\ref{tab:eval-static-resources}). Its peak memory is 0.99--1.68 times the corresponding \texttt{cargo check} memory. These measurements establish that the execution overhead of the static analysis stage is moderate. Audit time is an additional component, included in the workflow comparison in RQ2.

\emph{Case study: the Moros keyboard lock.}
Figure~\ref{fig:eval-case-study} illustrates how our workflow turns a static candidate into an agent assessment, using a confirmed defect involving \texttt{KEYBOARD}. Static analysis reports an interrupt self-dependency on this lock and identifies two source locations: the holding site in \texttt{set\_keyboard} (A) and an acquisition in the ISR-reachable callback (B). Together with the lock identity and lock/IRQ context, these locations provide entry points for the agent's source inspection.

The agent's recorded explanation and source citations document how these clues guide its checks. At the reported locations in \texttt{src/sys/keyboard.rs}, it checks that both operations access \texttt{KEYBOARD} and that the assignment in \texttt{set\_keyboard} acquires the lock without masking interrupts. To determine whether this holding site is reachable during normal execution, it examines the caller and command dispatcher shown in panel~(c). These connect the keyboard set shell command to the reported assignment, supporting a runtime path beyond the initialization call.

The report's ISR context also directs the agent to check how the callback becomes reachable from a hardware interrupt. These checks establish that the callback is registered for IRQ1, that registration unmasks the IRQ line, and that the hardware entry dispatches to the reported callback. The cited initialization code in \texttt{src/lib.rs:23--30} provides the surrounding setup context. Thus, the agent supplements the reported dependency with evidence about runtime reachability, interrupt registration, and dispatch.

The agent combines these findings to assign \texttt{high} confidence. If IRQ1 preempts the assignment on the same CPU while its temporary guard holds \texttt{KEYBOARD}, the callback spins on the same lock. The interrupted holder cannot resume to release the lock until the callback returns, producing a deadlock. This case illustrates the division of work: static analysis localizes the locking interaction, and the agent uses those locations to investigate the conditions needed to support its feasibility.

\begin{figure}[!tbp]
\centering
\resizebox{\linewidth}{!}{\begingroup%
\renewcommand{\ttdefault}{zi4}%
\definecolor{caseInk}{HTML}{253443}%
\definecolor{caseBlue}{HTML}{246887}%
\definecolor{caseRust}{HTML}{A24B23}%
\definecolor{caseSyntax}{HTML}{365E78}%
\definecolor{caseMuted}{HTML}{67737C}%
\definecolor{caseRule}{HTML}{BCC6CE}%
\begin{tikzpicture}[x=1cm,y=-.88cm,text=black,
  every node/.style={inner sep=0pt,outer sep=0pt,
    font=\normalfont\rmfamily\fontsize{8.5}{10}\selectfont\mdseries},
  heading/.style={anchor=west,
    font=\normalfont\rmfamily\fontsize{8.5}{10}\selectfont\bfseries},
  info/.style={anchor=west,text=black,
    font=\normalfont\rmfamily\fontsize{7.5}{8.5}\selectfont\bfseries},
  report/.style={anchor=north west,align=left,
    font=\normalfont\rmfamily\fontsize{7.5}{8.5}\selectfont\mdseries},
  code/.style={anchor=west,font=\ttfamily\fontsize{7.1}{8}\selectfont},
  line number/.style={anchor=east,text=caseMuted,font=\fontsize{6.3}{7}\selectfont},
  ref/.style={anchor=west,text=caseMuted,font=\ttfamily\fontsize{6.5}{7.5}\selectfont},
  badge/.style={circle,minimum size=3.5mm,text=white,
    font=\normalfont\rmfamily\fontsize{8.5}{10}\selectfont\bfseries},
  flow/.style={draw=caseBlue,line width=.8pt,-{Stealth[length=1.7mm,width=1.2mm]}},
  flow label/.style={anchor=west,align=left,
    font=\normalfont\rmfamily\fontsize{7.5}{8.5}\selectfont\mdseries}
]
\path[use as bounding box] (0,0) rectangle (14,11.07);

\path[draw=caseRule,line width=.55pt,rounded corners=1mm,fill=white]
  (0,0) rectangle (14,5.45);
\node[heading] at (.23,.34) {(a) Source code};
\node[ref,anchor=east] at (13.77,.34) {sys/keyboard.rs};
\draw[caseRule,line width=.4pt] (.23,.70) -- (13.77,.70);
\fill[caseBlue!4] (.02,.76) rectangle (13.98,1.34);

\node[line number] at (0.42,1.05) {12};
\node[code] at (0.61,1.05) {\hspace*{0.0em}\textcolor{caseSyntax}{pub} \textcolor{caseSyntax}{static} KEYBOARD: Mutex\textless{}Option\textless{}KeyboardLayout\textgreater{}\textgreater{} = Mutex::new(None);};

\node[info,anchor=east]
  at (13.75,1.05) {shared spin lock};
\draw[caseRule,line width=.4pt] (7,1.50) -- (7,5.22);
\node[info] at (.23,1.68) {Runtime shell command};
\node[info] at (7.23,1.68) {Keyboard interrupt callback};

\fill[caseBlue!10] (.53,3.00) rectangle (6.54,3.75);
\fill[caseRust!10] (7.53,2.26) rectangle (13.54,3.01);
\node[badge,fill=caseBlue] at (6.73,3.37) {A};
\node[badge,fill=caseRust] at (13.73,2.64) {B};

\node[line number] at (0.42,2.08) {63};
\node[code] at (0.61,2.08) {\hspace*{0.0em}\textcolor{caseSyntax}{pub} \textcolor{caseSyntax}{fn} set\_keyboard(layout: \&\textcolor{caseSyntax}{str}) -\textgreater{} \textcolor{caseSyntax}{bool} \{};
\node[line number] at (0.42,2.45) {64};
\node[code] at (0.61,2.45) {\hspace*{2.0em}\textcolor{caseSyntax}{if} \textcolor{caseSyntax}{let} Some(keyboard) =};
\node[line number] at (0.42,2.82) {};
\node[code] at (0.61,2.82) {\hspace*{4.0em}KeyboardLayout::from(layout) \{};
\node[line number] at (0.42,3.19) {65};
\node[code] at (0.61,3.19) {\hspace*{4.0em}*KEYBOARD.lock() =};
\node[line number] at (0.42,3.56) {};
\node[code] at (0.61,3.56) {\hspace*{6.0em}Some(keyboard);};
\node[line number] at (0.42,3.93) {66};
\node[code] at (0.61,3.93) {\hspace*{4.0em}\textcolor{caseSyntax}{true}};
\node[line number] at (0.42,4.30) {};
\node[code] at (0.61,4.30) {\hspace*{2.0em}\} \textcolor{caseSyntax}{else} \{ \textcolor{caseSyntax}{false} \}};
\node[line number] at (0.42,4.67) {70};
\node[code] at (0.61,4.67) {\hspace*{0.0em}\}};
\node[line number] at (7.42,2.08) {94};
\node[code] at (7.61,2.08) {\hspace*{0.0em}\textcolor{caseSyntax}{fn} interrupt\_handler() \{};
\node[line number] at (7.42,2.45) {95};
\node[code] at (7.61,2.45) {\hspace*{2.0em}\textcolor{caseSyntax}{if} \textcolor{caseSyntax}{let} Some(\textcolor{caseSyntax}{ref} \textcolor{caseSyntax}{mut} keyboard) =};
\node[line number] at (7.42,2.82) {};
\node[code] at (7.61,2.82) {\hspace*{4.0em}*KEYBOARD.lock() \{};
\node[line number] at (7.42,3.19) {96};
\node[code] at (7.61,3.19) {\hspace*{4.0em}\textcolor{caseSyntax}{let} scancode = read\_scancode();};
\node[line number] at (7.42,3.56) {};
\node[code] at (7.61,3.56) {\hspace*{4.0em}\textcolor{caseMuted}{// ... process key event ...}};
\node[line number] at (7.42,3.93) {};
\node[code] at (7.61,3.93) {\hspace*{2.0em}\}};
\node[line number] at (7.42,4.30) {};
\node[code] at (7.61,4.30) {\hspace*{0.0em}\}};

\node[anchor=west,font=\normalfont\rmfamily\fontsize{7.5}{8.5}\selectfont]
  at (.23,5.10) {\textcolor{caseBlue}{\textbf{A}}: temporary guard holds the lock during assignment.};
\node[anchor=west,font=\normalfont\rmfamily\fontsize{7.5}{8.5}\selectfont]
  at (7.23,5.10) {\textcolor{caseRust}{\textbf{B}}: the callback attempts to acquire the same lock.};

\draw[flow] (2.55,5.45) -- (2.55,6.30);
\node[flow label] at (2.75,5.87) {Extract lock\\dependency};
\draw[flow,densely dashed] (10.50,5.45) -- (10.50,6.30);
\node[flow label] at (10.70,5.87) {Inspect runtime\\and IRQ paths};

\path[draw=caseRule,line width=.55pt,rounded corners=1mm,fill=white]
  (0,6.30) rectangle (6.25,11.05);
\node[heading] at (.23,6.60) {(b) Static analysis};
\draw[caseRule,line width=.4pt] (.23,6.94) -- (6.02,6.94);
\node[report,text width=5.79cm] at (.23,7.17) {
  \raggedright
  \textbf{report\_id:} moros-ed37afc3cc91e6ce\\[.8mm]
  \textbf{lock instance:} KEYBOARD\\[.8mm]
  \textbf{dependency:} KEYBOARD \textrightarrow{} KEYBOARD\\
  \textbf{edge type:} interrupt\\[.8mm]
  \textbf{holding site:} \textcolor{caseBlue}{\textbf{A}} sys/keyboard.rs:65\\[.8mm]
  \textbf{acquisition site:} \textcolor{caseRust}{\textbf{B}} sys/keyboard.rs:95\\[.8mm]
  \textbf{lock / IRQ context:}\\[.4mm]
  \textcolor{caseBlue}{\textbf{A}} may hold KEYBOARD;
  \textcolor{caseRust}{\textbf{B}} is ISR-reachable.
  Runtime IRQ feasibility requires source inspection.
};

\draw[flow] (6.25,9.11) -- (7.50,9.11);
\node[flow label,anchor=center] at (6.875,8.86) {Report};

\path[draw=caseRule,line width=.55pt,rounded corners=1mm,fill=white]
  (7.50,6.30) rectangle (14,11.05);
\node[heading] at (7.73,6.60) {(c) Agent assessment};
\draw[caseRule,line width=.4pt] (7.73,6.94) -- (13.77,6.94);
\node[report,text width=6.04cm] at (7.73,7.17) {
  \raggedright
  \textbf{report\_id:} moros-ed37afc3cc91e6ce\\[.8mm]
  \textbf{confidence:} high\\[.8mm]
  \textbf{reason:}\par\vspace{.5mm}
  \hangindent=1em\hangafter=1
  1. The shell command reaches \textcolor{caseBlue}{\textbf{A}} without masking IRQs
  (usr/shell.rs:547; usr/keyboard.rs:17).\par\vspace{.6mm}
  \hangindent=1em\hangafter=1
  2. IRQ1 is registered/unmasked and dispatches to \textcolor{caseRust}{\textbf{B}}
  (sys/keyboard.rs:74; sys/idt.rs:71--86, 300--307).\par\vspace{.6mm}
  \hangindent=1em\hangafter=1
  3. If IRQ1 preempts \textcolor{caseBlue}{\textbf{A}} on the same CPU while
  \textcolor{caseBlue}{\textbf{A}} holds KEYBOARD, \textcolor{caseRust}{\textbf{B}} spins
  and \textcolor{caseBlue}{\textbf{A}} cannot resume to release it:
  interrupt-induced self-deadlock.\par
};
\end{tikzpicture}%
\endgroup%
}
\input{figures/evaluation-case-study/caption.tex}
\Description{Source acquisition A is in set_keyboard; B is in the keyboard interrupt callback. Static analysis identifies a potential interrupt self-cycle on the same lock. The candidate and source both feed the agent assessment. The structured static report gives the identifier, lock, dependency, sites, and context. The agent output contains report_id, confidence, and reason, including runtime and IRQ checks and a textual explanation of the same-CPU circular wait.}
\label{fig:eval-case-study}
\end{figure}

\emph{Answer to RQ1.}
The complete workflow identifies 26 true cases associated with nine confirmed defects across four Rust operating systems, and assigns high confidence to 25 of those cases. The keyboard example illustrates how interrupt-aware candidate generation and source-based assessment connect to an actual defect.

\subsection{RQ2: Baseline Comparison and Stage Contributions}
\label{sec:eval-rq2}

\emph{Comparison design.}
We compare the complete workflow (\emph{Full}) with LockBud~\cite{qin2024understanding}, a state-of-the-art static Rust deadlock detector, and \emph{Agent-only}, which searches the same repository without static reports. LockBud does not model interrupt behavior and can only detect deadlocks arising from ordinary multithreaded execution. We therefore use it as an external baseline to provide supporting evidence for the necessity of modeling interrupts. The agents in Full and Agent-only use the same model and operate under the same access restrictions, but their tasks differ: one assesses supplied candidates, while the other must first discover them.

\begin{table}[!htbp]
\centering
\caption{Case-level results for the full pipeline, LockBud, and agent-only audit.}

\begingroup
\fontsize{8.8}{10.5}\selectfont
\renewcommand{\arraystretch}{1.05}
\setlength{\extrarowheight}{0pt}
\setlength{\aboverulesep}{1.5pt}
\setlength{\belowrulesep}{2pt}
\setlength{\tabcolsep}{2.4pt}
\noindent
\begin{minipage}{\linewidth}
\begin{tabular*}{\linewidth}{@{\extracolsep{\fill}}lrrrrrr@{}}
\toprule
 & \multicolumn{2}{c}{Full} & \multicolumn{2}{c}{LockBud} & \multicolumn{2}{c}{Agent-only} \\
\cmidrule(lr){2-3}\cmidrule(lr){4-5}\cmidrule(l){6-7}
Target & Cases & True & Cases & True & Cases & True \\
\midrule
\textsc{Asterinas} & 9 & \cellcolor{black!8}9 & 22 & \cellcolor{black!8}0 & 2 & \multicolumn{1}{>{\columncolor{black!8}[\tabcolsep][0pt]}r@{}}{0} \\
\textsc{DragonOS} & 14 & \cellcolor{black!8}10 & N/A & \cellcolor{black!8}N/A & 1 & \multicolumn{1}{>{\columncolor{black!8}[\tabcolsep][0pt]}r@{}}{0} \\
\textsc{Motor-OS} & 2 & \cellcolor{black!8}2 & 0 & \cellcolor{black!8}0 & 0 & \multicolumn{1}{>{\columncolor{black!8}[\tabcolsep][0pt]}r@{}}{0} \\
\textsc{Moros} & 4 & \cellcolor{black!8}4 & 0 & \cellcolor{black!8}0 & 2 & \multicolumn{1}{>{\columncolor{black!8}[\tabcolsep][0pt]}r@{}}{2} \\
\midrule
\textbf{All four projects} & \textbf{29} & \cellcolor{black!8}\textbf{25} & N/A & \cellcolor{black!8}N/A & \textbf{5} & \multicolumn{1}{>{\columncolor{black!8}[\tabcolsep][0pt]}r@{}}{\textbf{2}} \\
Three shared projects & 15 & \cellcolor{black!8}15 & 22 & \cellcolor{black!8}0 & 4 & \multicolumn{1}{>{\columncolor{black!8}[\tabcolsep][0pt]}r@{}}{2} \\
\bottomrule
\end{tabular*}
\par\vspace{3pt}
{\fontsize{8}{10}\selectfont\noindent Full lists its cases labeled high-confidence by the agent auditor. Agent-only's two true cases correspond to bugs among the nine distinct defects found by Full. N/A denotes \textsc{DragonOS}'s incompatibility with LockBud because of its Rust toolchain version and custom lock implementations. The shared-project subtotal includes \textsc{Asterinas}, \textsc{Motor-OS}, and \textsc{Moros}. \par}
\end{minipage}
\endgroup

\label{tab:eval-baselines}
\end{table}

\emph{Detection relative to baselines.}
Table~\ref{tab:eval-baselines} compares Full's high-confidence cases with the reports emitted by the two baselines. On the three projects supported by all methods, Full marks 15 true cases as high confidence, whereas LockBud reports no true cases. LockBud cannot be evaluated on \textsc{DragonOS} because of toolchain and custom-lock incompatibilities.

Across all four projects, Full's high-confidence subset contains 25 true cases, compared with two among Agent-only's five reports. Both true cases reported by Agent-only correspond to defects already included among the nine distinct bugs detected by our workflow. Supplying static candidates therefore helps the agent identify substantially more true cases in this evaluation. The results demonstrate a benefit of the combined workflow.

Since every high-confidence case originates from a candidate produced by the static-analysis stage, these results indicate that modeling interrupt behavior is necessary for static deadlock detection in operating system projects.

\begin{table}[!htbp]
\centering
\caption{Candidate quality before and after agent auditing.}

\begingroup
\fontsize{8.8}{10.5}\selectfont
\renewcommand{\arraystretch}{1.05}
\setlength{\extrarowheight}{0pt}
\setlength{\aboverulesep}{1.5pt}
\setlength{\belowrulesep}{2pt}
\setlength{\tabcolsep}{2.8pt}
\noindent
\begin{minipage}{\linewidth}
\begin{tabular*}{\linewidth}{@{\extracolsep{\fill}}lrrrrrr@{}}
\toprule
 & \multicolumn{2}{c}{True-case proportion} & \multicolumn{2}{c}{True-case recall} & \multicolumn{2}{c}{False-case filter} \\
\cmidrule(lr){2-3}\cmidrule(lr){4-5}\cmidrule(l){6-7}
Target & \shortstack[r]{Before audit} & \shortstack[r]{After audit} & \shortstack[r]{True cases\\assigned high} & Recall rate & \shortstack[r]{False cases\\assigned low} & Filter rate \\
\midrule
\textsc{Asterinas} & 47.6\% & 100.0\% & 9/10 & 90.0\% & 11/11 & 100.0\% \\
\textsc{DragonOS} & 14.3\% & 71.4\% & 10/10 & 100.0\% & 56/60 & 93.3\% \\
\textsc{Motor-OS} & 100.0\% & 100.0\% & 2/2 & 100.0\% & 0/0 & N/A \\
\textsc{Moros} & 10.3\% & 100.0\% & 4/4 & 100.0\% & 35/35 & 100.0\% \\
\midrule
\textbf{Total} & \textbf{19.7\%} & \textbf{86.2\%} & \textbf{25/26} & \textbf{96.2\%} & \textbf{102/106} & \textbf{96.2\%} \\
\bottomrule
\end{tabular*}
\par\vspace{3pt}
{\fontsize{8}{10}\selectfont\noindent Before audit uses all candidates generated by the static analysis stage; after audit uses the high-confidence cases reported by the agent. Recall rate is measured within the known true candidates rather than over all repository bugs. \par}
\end{minipage}
\endgroup

\label{tab:eval-audit-quality}
\end{table}

\emph{Contribution of agent auditing.}
We compare the true-case proportion among all static candidates with precision in the high-confidence subset, and report the fraction of detected true candidates assigned high confidence. Table~\ref{tab:eval-audit-quality} shows that the proportion of false positives decreases from 80.3\% (106/132) in the static-analysis results to 13.8\% (4/29) in the high-confidence subset after agent auditing, indicating that agent auditing effectively filters out false positives. The high-confidence subset contains 25/26 (96.2\%) of the true cases. This recall is relative to the detected candidate set, not the ground-truth recall over all defects in a repository. The effect is particularly visible on \textsc{DragonOS} and \textsc{Moros}. \textsc{Motor-OS} has no false candidates, so auditing provides supporting explanations without improving its already perfect candidate precision.

A human auditor who starts with the 29 high-confidence cases examines a subset 78.0\% smaller than the original 132 candidates, while encountering 25 of the 26 true cases. The remaining true case is still present among the low-confidence reports and requires labor, so cases labeled low should not simply be excluded from manual review. However, the result still indicates clear prioritization for human auditors.

\emph{Runtime and token cost.}
In terms of static-analysis time (Figure~\ref{fig:eval-costs}), Full's static stage is comparable to LockBud, although LockBud reports no true deadlocks. In terms of end-to-end execution time, Full and Agent-only each perform better on different projects. Full requires the agent to audit every report, so the time required is positively correlated with the number of reports. \textsc{DragonOS} produces 70 reports to audit, resulting in a substantially longer runtime.

\begin{figure}[!tbp]
\centering
\resizebox{\linewidth}{!}{\definecolor{evalStatic}{HTML}{295B7B}
\definecolor{evalAudit}{HTML}{8BB7D0}
\definecolor{evalFull}{HTML}{3D7A9F}
\definecolor{evalAgent}{HTML}{C08232}
\definecolor{evalAgentOutput}{HTML}{E6BE87}
\definecolor{evalLockbud}{HTML}{8D969D}
\definecolor{evalGrid}{HTML}{E3E6E8}

\begin{tikzpicture}
\begin{axis}[at={(0,0.00cm)},anchor=north west,
width=136mm,height=40mm,
xmin=-.6,xmax=4.2,ymin=0,ymax=920,ytick={0,200,400,600,800},
xtick={0,1.2,2.4,3.6},xticklabels={Asterinas,DragonOS,Motor-OS,Moros},
axis x line*=bottom,axis y line*=left,axis line style={black!40},
tick style={black!35},tick align=outside,
yticklabel style={font=\fontsize{8.5}{10}\selectfont},
xticklabel style={font=\fontsize{8.5}{10}\selectfont},
scaled ticks=false,ymajorgrids=true,grid style={evalGrid,line width=.3pt},
enlarge x limits=false,clip=false,
label style={font=\fontsize{9}{11}\selectfont},
title style={at={(0,1.34)},anchor=west,font=\normalfont\rmfamily\fontsize{9}{11}\selectfont\bfseries},
legend style={draw=none,fill=none,font=\fontsize{8}{9.5}\selectfont,
at={(1,1.08)},anchor=south east,/tikz/every even column/.append style={column sep=5pt}},
legend cell align=left,
ylabel={Time (s)},title={(a) Execution time},legend columns=4]

\addlegendimage{area legend,fill=evalStatic,draw=evalStatic!85!black}\addlegendentry{Full: static}
\addlegendimage{area legend,fill=evalAudit,draw=evalAudit!85!black}\addlegendentry{Full: audit}
\addlegendimage{area legend,fill=evalAgent,draw=evalAgent!85!black,postaction={pattern=north east lines,pattern color=evalAgent!70!black}}\addlegendentry{Agent-only}
\addlegendimage{area legend,fill=evalLockbud,draw=evalLockbud!85!black}\addlegendentry{LockBud}
\path[fill=evalStatic,draw=evalStatic!85!black,line width=.3pt] (axis cs:-0.375,0.000) rectangle (axis cs:-0.165,110.200);

\path[fill=evalAudit,draw=evalAudit!85!black,line width=.3pt] (axis cs:-0.375,110.200) rectangle (axis cs:-0.165,414.300);

\node[anchor=south,font=\fontsize{8}{9}\selectfont,inner sep=1.5pt] at (axis cs:-0.270,414.300) {414.3};

\path[fill=evalAgent,draw=evalAgent!85!black,line width=.3pt,postaction={pattern=north east lines,pattern color=evalAgent!70!black}] (axis cs:-0.105,0.000) rectangle (axis cs:0.105,372.100);

\node[anchor=south,font=\fontsize{8}{9}\selectfont,inner sep=1.5pt] at (axis cs:0.000,372.100) {372.1};

\path[fill=evalLockbud,draw=evalLockbud!85!black,line width=.3pt] (axis cs:0.165,0.000) rectangle (axis cs:0.375,100.800);

\node[anchor=south,font=\fontsize{8}{9}\selectfont,inner sep=1.5pt] at (axis cs:0.270,100.800) {100.8};

\path[fill=evalStatic,draw=evalStatic!85!black,line width=.3pt] (axis cs:0.825,0.000) rectangle (axis cs:1.035,102.100);

\path[fill=evalAudit,draw=evalAudit!85!black,line width=.3pt] (axis cs:0.825,102.100) rectangle (axis cs:1.035,549.000);

\node[anchor=south,font=\fontsize{8}{9}\selectfont,inner sep=1.5pt] at (axis cs:0.930,549.000) {549.0};

\path[fill=evalAgent,draw=evalAgent!85!black,line width=.3pt,postaction={pattern=north east lines,pattern color=evalAgent!70!black}] (axis cs:1.095,0.000) rectangle (axis cs:1.305,261.600);

\node[anchor=south,font=\fontsize{8}{9}\selectfont,inner sep=1.5pt] at (axis cs:1.200,261.600) {261.6};

\node[anchor=south,font=\fontsize{8}{9}\selectfont,text=black!65,align=center] at (axis cs:1.470,15) {N/A};
\path[fill=evalStatic,draw=evalStatic!85!black,line width=.3pt] (axis cs:2.025,0.000) rectangle (axis cs:2.235,41.300);

\path[fill=evalAudit,draw=evalAudit!85!black,line width=.3pt] (axis cs:2.025,41.300) rectangle (axis cs:2.235,237.600);

\node[anchor=south,font=\fontsize{8}{9}\selectfont,inner sep=1.5pt] at (axis cs:2.130,237.600) {237.6};

\path[fill=evalAgent,draw=evalAgent!85!black,line width=.3pt,postaction={pattern=north east lines,pattern color=evalAgent!70!black}] (axis cs:2.295,0.000) rectangle (axis cs:2.505,794.400);

\node[anchor=south,font=\fontsize{8}{9}\selectfont,inner sep=1.5pt] at (axis cs:2.400,794.400) {794.4};

\path[fill=evalLockbud,draw=evalLockbud!85!black,line width=.3pt] (axis cs:2.565,0.000) rectangle (axis cs:2.775,30.300);

\node[anchor=south,font=\fontsize{8}{9}\selectfont,inner sep=1.5pt] at (axis cs:2.670,30.300) {30.3};

\path[fill=evalStatic,draw=evalStatic!85!black,line width=.3pt] (axis cs:3.225,0.000) rectangle (axis cs:3.435,19.500);

\path[fill=evalAudit,draw=evalAudit!85!black,line width=.3pt] (axis cs:3.225,19.500) rectangle (axis cs:3.435,206.100);

\node[anchor=south,font=\fontsize{8}{9}\selectfont,inner sep=1.5pt] at (axis cs:3.330,206.100) {206.1};

\path[fill=evalAgent,draw=evalAgent!85!black,line width=.3pt,postaction={pattern=north east lines,pattern color=evalAgent!70!black}] (axis cs:3.495,0.000) rectangle (axis cs:3.705,383.700);

\node[anchor=south,font=\fontsize{8}{9}\selectfont,inner sep=1.5pt] at (axis cs:3.600,383.700) {383.7};

\path[fill=evalLockbud,draw=evalLockbud!85!black,line width=.3pt] (axis cs:3.765,0.000) rectangle (axis cs:3.975,55.300);

\node[anchor=south,font=\fontsize{8}{9}\selectfont,inner sep=1.5pt] at (axis cs:3.870,55.300) {55.3};

\end{axis}
\begin{axis}[at={(0,-4.40cm)},anchor=north west,
width=136mm,height=40mm,
xmin=-.6,xmax=4.2,ymin=0,ymax=26,ytick={0,5,10,15,20,25},
xtick={0,1.2,2.4,3.6},xticklabels={Asterinas,DragonOS,Motor-OS,Moros},
axis x line*=bottom,axis y line*=left,axis line style={black!40},
tick style={black!35},tick align=outside,
yticklabel style={font=\fontsize{8.5}{10}\selectfont},
xticklabel style={font=\fontsize{8.5}{10}\selectfont},
scaled ticks=false,ymajorgrids=true,grid style={evalGrid,line width=.3pt},
enlarge x limits=false,clip=false,
label style={font=\fontsize{9}{11}\selectfont},
title style={at={(0,1.34)},anchor=west,font=\normalfont\rmfamily\fontsize{9}{11}\selectfont\bfseries},
legend style={draw=none,fill=none,font=\fontsize{8}{9.5}\selectfont,
at={(1,1.08)},anchor=south east,/tikz/every even column/.append style={column sep=5pt}},
legend cell align=left,
ylabel={Reference cost (USD)},title={(b) Token cost},legend columns=4]

\addlegendimage{area legend,fill=evalFull,draw=evalFull!85!black}\addlegendentry{Full: input}
\addlegendimage{area legend,fill=evalAudit,draw=evalAudit!85!black}\addlegendentry{Full: output}
\addlegendimage{area legend,fill=evalAgent,draw=evalAgent!85!black,postaction={pattern=north east lines,pattern color=evalAgent!70!black}}\addlegendentry{Agent-only: input}
\addlegendimage{area legend,fill=evalAgentOutput,draw=evalAgentOutput!85!black,postaction={pattern=north east lines,pattern color=evalAgentOutput!70!black}}\addlegendentry{Agent-only: output}
\path[fill=evalFull,draw=evalFull!85!black,line width=.3pt] (axis cs:-0.265,0.000) rectangle (axis cs:-0.055,2.491);

\path[fill=evalAudit,draw=evalAudit!85!black,line width=.3pt] (axis cs:-0.265,2.491) rectangle (axis cs:-0.055,3.831);

\node[anchor=south,font=\fontsize{8}{9}\selectfont,inner sep=1.5pt] at (axis cs:-0.160,3.831) {\$3.83};

\path[fill=evalAgent,draw=evalAgent!85!black,line width=.3pt,postaction={pattern=north east lines,pattern color=evalAgent!70!black}] (axis cs:0.055,0.000) rectangle (axis cs:0.265,7.565);

\path[fill=evalAgentOutput,draw=evalAgentOutput!85!black,line width=.3pt,postaction={pattern=north east lines,pattern color=evalAgentOutput!70!black}] (axis cs:0.055,7.565) rectangle (axis cs:0.265,8.645);

\node[anchor=south,font=\fontsize{8}{9}\selectfont,inner sep=1.5pt] at (axis cs:0.160,8.645) {\$8.64};

\path[fill=evalFull,draw=evalFull!85!black,line width=.3pt] (axis cs:0.935,0.000) rectangle (axis cs:1.145,3.182);

\path[fill=evalAudit,draw=evalAudit!85!black,line width=.3pt] (axis cs:0.935,3.182) rectangle (axis cs:1.145,5.142);

\node[anchor=south,font=\fontsize{8}{9}\selectfont,inner sep=1.5pt] at (axis cs:1.040,5.142) {\$5.14};

\path[fill=evalAgent,draw=evalAgent!85!black,line width=.3pt,postaction={pattern=north east lines,pattern color=evalAgent!70!black}] (axis cs:1.255,0.000) rectangle (axis cs:1.465,3.995);

\path[fill=evalAgentOutput,draw=evalAgentOutput!85!black,line width=.3pt,postaction={pattern=north east lines,pattern color=evalAgentOutput!70!black}] (axis cs:1.255,3.995) rectangle (axis cs:1.465,4.815);

\node[anchor=south,font=\fontsize{8}{9}\selectfont,inner sep=1.5pt] at (axis cs:1.360,4.815) {\$4.81};

\path[fill=evalFull,draw=evalFull!85!black,line width=.3pt] (axis cs:2.135,0.000) rectangle (axis cs:2.345,1.019);

\path[fill=evalAudit,draw=evalAudit!85!black,line width=.3pt] (axis cs:2.135,1.019) rectangle (axis cs:2.345,1.719);

\node[anchor=south,font=\fontsize{8}{9}\selectfont,inner sep=1.5pt] at (axis cs:2.240,1.719) {\$1.72};

\path[fill=evalAgent,draw=evalAgent!85!black,line width=.3pt,postaction={pattern=north east lines,pattern color=evalAgent!70!black}] (axis cs:2.455,0.000) rectangle (axis cs:2.665,19.825);

\path[fill=evalAgentOutput,draw=evalAgentOutput!85!black,line width=.3pt,postaction={pattern=north east lines,pattern color=evalAgentOutput!70!black}] (axis cs:2.455,19.825) rectangle (axis cs:2.665,22.265);

\node[anchor=south,font=\fontsize{8}{9}\selectfont,inner sep=1.5pt] at (axis cs:2.560,22.265) {\$22.26};

\path[fill=evalFull,draw=evalFull!85!black,line width=.3pt] (axis cs:3.335,0.000) rectangle (axis cs:3.545,0.959);

\path[fill=evalAudit,draw=evalAudit!85!black,line width=.3pt] (axis cs:3.335,0.959) rectangle (axis cs:3.545,1.699);

\node[anchor=south,font=\fontsize{8}{9}\selectfont,inner sep=1.5pt] at (axis cs:3.440,1.699) {\$1.70};

\path[fill=evalAgent,draw=evalAgent!85!black,line width=.3pt,postaction={pattern=north east lines,pattern color=evalAgent!70!black}] (axis cs:3.655,0.000) rectangle (axis cs:3.865,3.920);

\path[fill=evalAgentOutput,draw=evalAgentOutput!85!black,line width=.3pt,postaction={pattern=north east lines,pattern color=evalAgentOutput!70!black}] (axis cs:3.655,3.920) rectangle (axis cs:3.865,5.400);

\node[anchor=south,font=\fontsize{8}{9}\selectfont,inner sep=1.5pt] at (axis cs:3.760,5.400) {\$5.40};

\end{axis}
\end{tikzpicture}}
\input{figures/evaluation-costs/caption.tex}
\Description{Two panels compare execution time and reference token cost for four projects. Token-cost bars stack input and output charges, with total USD prices above the bars. Full has lower token cost on Asterinas, Motor-OS, and Moros, but slightly higher cost on DragonOS.}
\label{fig:eval-costs}
\end{figure}

Full has substantially lower token costs than Agent-only on all projects except \textsc{DragonOS}, where the larger number of reports makes its cost slightly higher: higher output charges outweigh input savings (Table~\ref{tab:eval-token-details}). In particular, the diagnostic-guided agent uses fewer input tokens on every project. Together with the detection results, this supports the value of static diagnostics as entry points for source inspection: static-analysis results give the agent a clear focus, guide its investigation, and reduce auditing costs in most cases.

\begin{table}[!htbp]
\centering
\caption{Token consumption (thousands of tokens) and reference cost (USD) for the two agent workflows.}

\begingroup
\fontsize{8.5}{10.5}\selectfont
\renewcommand{\arraystretch}{1.05}
\setlength{\extrarowheight}{0pt}
\setlength{\aboverulesep}{1.5pt}
\setlength{\belowrulesep}{2pt}
\setlength{\tabcolsep}{2.0pt}
\noindent
\begin{minipage}{\linewidth}
\begin{tabular*}{\linewidth}{@{\extracolsep{\fill}}lrrrrrrrr@{}}
\toprule
 & \multicolumn{4}{c}{Full: diagnostic-guided audit} & \multicolumn{4}{c}{Agent-only} \\
\cmidrule(lr){2-5}\cmidrule(l){6-9}
Target & Input & Cached & Output & Ref. cost(USD) & Input & Cached & Output & Ref. cost(USD) \\
\midrule
\textsc{Asterinas} & 5,184 & 5,068 & 67 & \cellcolor{black!8}3.83 & 16,914 & 16,692 & 54 & \multicolumn{1}{>{\columncolor{black!8}[\tabcolsep][0pt]}r@{}}{8.64} \\
\textsc{DragonOS} & 6,417 & 6,246 & 98 & \cellcolor{black!8}5.14 & 8,628 & 8,477 & 41 & \multicolumn{1}{>{\columncolor{black!8}[\tabcolsep][0pt]}r@{}}{4.81} \\
\textsc{Motor-OS} & 1,999 & 1,938 & 35 & \cellcolor{black!8}1.72 & 46,718 & 46,402 & 122 & \multicolumn{1}{>{\columncolor{black!8}[\tabcolsep][0pt]}r@{}}{22.26} \\
\textsc{Moros} & 1,786 & 1,718 & 37 & \cellcolor{black!8}1.70 & 8,559 & 8,421 & 74 & \multicolumn{1}{>{\columncolor{black!8}[\tabcolsep][0pt]}r@{}}{5.40} \\
\midrule
\textbf{Total} & \textbf{15,386} & \textbf{14,970} & \textbf{237} & \cellcolor{black!8}\textbf{12.39} & \textbf{80,819} & \textbf{79,992} & \textbf{291} & \multicolumn{1}{>{\columncolor{black!8}[\tabcolsep][0pt]}r@{}}{\textbf{41.12}} \\
\bottomrule
\end{tabular*}
\par\vspace{3pt}
{\fontsize{8}{10}\selectfont\noindent Token counts are in K(thousand). Cached means cached input(included in Input). Reference cost uses \$4.00 for uncached input, \$0.40 for cached input, and \$20.00 for output, per million tokens. \par}
\end{minipage}
\endgroup

\label{tab:eval-token-details}
\end{table}

\emph{Answer to RQ2.}
The combination finds more true cases than either baseline in the supported comparisons. Static diagnostics accompany a reduction in agent input consumption, while the internal before/after comparison shows that auditing concentrates true cases in the high-confidence subset. These complementary contributions support the complete workflow, with project-dependent runtime and output costs.

\subsection{RQ3: Annotation Effort}
\label{sec:eval-rq3}

We count the five kinds of semantic annotations to measure the manual configuration surface, and the recovered lock and guard instances to indicate how much information the analyzer obtains from this configuration.

\begin{table}[!htbp]
\centering
\caption{Semantic annotation counts and recovered lock and guard instances for each project.}

\begingroup
\fontsize{8.5}{10.5}\selectfont
\renewcommand{\arraystretch}{1.05}
\setlength{\extrarowheight}{0pt}
\setlength{\aboverulesep}{1.5pt}
\setlength{\belowrulesep}{2pt}
\setlength{\tabcolsep}{2.0pt}
\noindent
\begin{minipage}{\linewidth}
\begin{tabular*}{\linewidth}{@{\extracolsep{\fill}}lrrrrrrrr@{}}
\toprule
 & \multicolumn{3}{c}{Entry / operation markers} & \multicolumn{2}{c}{Type markers} &  & \multicolumn{2}{c}{Recovered instances} \\
\cmidrule(lr){2-4}\cmidrule(lr){5-6}\cmidrule(l){8-9}
Target & IsrEntry & IntrApi & LockOp & LockType & LockGuardType & \textbf{Total} & Locks & Guards \\
\midrule
\textsc{Asterinas} & 9 & 2 & 8 & 4 & 4 & \cellcolor{black!8}\textbf{27} & 283 & 1,472 \\
\textsc{DragonOS} & 2 & 4 & 12 & 4 & 4 & \cellcolor{black!8}\textbf{26} & 904 & 4,798 \\
\textsc{Motor-OS} & 1 & 3 & 1 & 1 & 1 & \cellcolor{black!8}\textbf{7} & 23 & 163 \\
\textsc{Moros} & 4 & 2 & 1 & 1 & 1 & \cellcolor{black!8}\textbf{9} & 38 & 132 \\
\midrule
\textbf{Total} & \textbf{16} & \textbf{11} & \textbf{22} & \textbf{10} & \textbf{10} & \cellcolor{black!8}\textbf{69} & \textbf{1,248} & \textbf{6,565} \\
\bottomrule
\end{tabular*}
\par\vspace{3pt}
{\fontsize{8}{10}\selectfont\noindent IsrEntry counts functions annotated as ISR entries. IntrApi, LockOp, LockType, and LockGuardType count annotated interrupt-control APIs, lock-acquisition APIs, lock structs, and lock guard structs, respectively.\par}
\end{minipage}
\endgroup

\label{tab:eval-annotations}
\end{table}

Table~\ref{tab:eval-annotations} shows that each project requires 7--27 annotations. The annotations concentrate on shared types, acquisition operations, explicit IRQ operations, and interrupt entries. Once the semantics of these abstractions are supplied, the analyzer applies them at their uses rather than requiring an annotation for every lock instance or acquisition.

\emph{Answer to RQ3.}
The measurements support a small, concentrated configuration surface for adapting the detector to different synchronization APIs, addressing Challenge~2.

\subsection{Discussion of Findings}
\label{sec:eval-discussion}

\emph{Effectiveness and costs.}
The nine confirmed defects establish the workflow's practical effectiveness across the evaluated kernels. The baseline comparison and token measurements support using static diagnostics to focus source inspection, while the candidate-quality comparison shows the benefit of agent assessments for prioritization. The smaller high-confidence subset provides an opportunity to prioritize human review.

\emph{Sources of false positives and false negatives.}
We attribute the false-positive cases generated by the static analysis stage to the following sources of imprecision. First, call-graph over-approximation can mark functions as ISR-reachable even when they are never executed in an actual interrupt context in the built configuration: conservative call edges and indirect calls inflate the ISR slice, so the analyzer may construct interrupt edges between normal paths and code that, in practice, only runs in thread context. Second, conservative pointer analysis can introduce imprecision. For example, the analyzer cannot precisely determine whether two method calls on one structure involve the same object instance, leading to some false positives. The main potential source of false negatives in static analysis is incomplete or incorrect annotations, which can cause the analyzer to miss relevant interrupts or locks.

In the agent auditing stage, the four false positives in our experiments arise because the LLM fails to account for the system's execution phase. The reported lock-order cycles do exist, but the associated code runs only during system startup, on a single CPU core with interrupts disabled, so these cycles cannot cause deadlocks in practice. The single false negative involves a possible panic while a lock is held, followed by an attempt to reacquire that lock in the panic handler. Here, the panic plays a preemptive role similar to that of an interrupt, but the LLM rejects the case because interrupts are disabled while the lock is held. Overall, agent auditing achieves high accuracy but remains susceptible to unusual corner cases and incomplete reasoning.

\emph{External Validation.}
Our tool is primarily designed to detect interrupt-induced deadlocks on mutually exclusive locks such as spinlocks and mutexes, a synchronization pattern that is widely used across operating-system kernels rather than being specific to a single Rust OS code base. The current results therefore provide some evidence of external validity within this scope: across kernels of different sizes, the required annotation effort remains manageable and the end-to-end runtime overhead stays scalable enough for practical offline analysis. Understanding primitive semantics and aligning the analyzer with a new project's Rust toolchain version remain separate engineering tasks beyond adding annotations, because the current implementation depends on compiler-internal interfaces.

\FloatBarrier

\section{Related Work}
\label{sec:related-work}

\subsection{Deadlock Detection Methods}
\label{sec:rw-deadlock}

Dynamic validators such as Linux's lockdep track lock-class dependencies and interrupt-context usage during execution to identify potential deadlocks~\cite{corbet2006lockdep,linuxlockdoc}. Although lockdep can infer unsafe combinations without reproducing the full deadlocking schedule, it still requires the component locking paths to execute. Rarely exercised kernel paths and hardware-dependent behavior therefore limit dynamic coverage. ThreadSanitizer and Eraser detect data races dynamically~\cite{serebryany2009tsan,savage1997eraser}. Static methods instead extract dependencies without requiring those paths to execute; RacerX uses flow-sensitive, interprocedural analysis to recover locking constraints and detect potential cycles~\cite{engler2003racerx}. DeadWait targets static deadlock detection in asynchronous C\# programs~\cite{santhiar2017deadwait}.

Other concurrency analyses use bug patterns~\cite{hong2013cobet}, explore schedules~\cite{fonseca2014ski} or inter-thread communication~\cite{gong2021snowboard}, and detect races statically~\cite{voung2007relay,dossche2024contextsensitiveoutlierbasedstaticanalysis} or interrupt-driven atomicity violations~\cite{li2022atomicity}. Kernel-specific static analyses account for execution contexts and the feasibility of lock cycles. DLOS combines summaries of lock usage, reachability-based cycle detection, and filtering of infeasible or nonconcurrent paths~\cite{bai2022dlos}. ArcherFish explicitly captures regular and interrupt-related dependencies in an interrupt-aware lock graph, with dedicated analyses to identify interrupt handlers and validate candidate cycles~\cite{ye2024archerfish}. These methods establish the importance of modeling kernel execution conditions when interpreting a lock dependency.

DLOS and ArcherFish target C kernels and recognize synchronization through explicit lock/unlock and interrupt-control operations. Applying their models to Rust OS code requires recovering equivalent effects from RAII guards: a lock guard may move across function boundaries, and destroying a guard may release a lock or restore a previous IRQ state. Our work retains prior kernel analyses' interrupt-aware perspective while making these effects available to dependency construction for project-specific Rust synchronization primitives.

\subsection{Rust System Programming}
\label{sec:rw-rust}

Rust is used in kernel extensions, embedded systems, and operating systems such as Theseus and Asterinas~\cite{li2024empirical,sharma2024rust,boos2020theseus,peng2025asterinas}. Empirical studies document memory and concurrency bugs beyond compiler guarantees~\cite{qin2020rust,xu2021memory}. Studies also examine unsafe-code usage and safety obligations~\cite{evans2020rust,astrauskas2020programmers,cui2024unsafe}, encapsulation~\cite{rao2024characterizing}, and annotation, auditing, and verification of unsafe abstractions~\cite{rao2025annotating,Rao2024SemiAutomatedVO}. Static bug-finding tools address memory and other safety errors~\cite{cui2023safedrop,bae2021rudra,li2021mirchecker}, while runtime defenses protect safe Rust from unsafe components~\cite{rivera2021keeping,liu2020securing}. Verification frameworks and semantic foundations support reasoning about Rust program safety~\cite{astrauskas2022prusti,jung2017rustbelt,matsushita2022rusthornbelt,lattuada2023verus,lattuada2024verus,gaher2024refinedrust,lehmann2023flux}; a survey summarizes this landscape~\cite{blanc2024surveying}.

Rust-specific deadlock analyses already reason about implicit lock release. Stuck-me-not uses MIR to detect double-lock bugs in blockchain software~\cite{ning2020stuckmenot}. Static Deadlock Detection for Rust Programs uses ownership- and lifetime-aware pointer and dependency analysis to identify double locks, conflicting lock orders, and condition-variable deadlocks~\cite{zhang2024rustdeadlock}. LockBud~\cite{qin2024understanding} recovers critical sections from lock-guard lifetimes in MIR, associates guards with the mutexes they protect, and uses live-guard information at acquisitions to identify double locks and conflicting lock orders.

These Rust analyses provide guard and lifetime reasoning, but do not model kernel interrupt preemption or recover project-specific IRQ semantics. Together with interrupt-aware C kernel analyses, they motivate our focus on jointly recovering guard-managed locking and interrupt state in Rust operating systems.

\section{Conclusion}
\label{sec:conclusion}

This paper presents an approach combining interrupt-aware static analysis with diagnostic-guided agent auditing for detecting deadlocks in Rust operating systems. Annotation-based metadata collection and joint lock and interrupt state analysis support the construction of an interrupt-aware LDG. The tool identifies candidate cycles and extracts their supporting evidence into structured reports, which guide source-based agent assessment. Each candidate receives a high/low confidence label with reasons grounded in source code. Across four real Rust OS projects, the candidate sets contain 26 true cases corresponding to nine maintainer-confirmed bugs. The high-confidence group achieves 86.2\% precision and contains 25 of these 26 true cases. Guided auditing uses 81.0\% fewer input tokens in total than agent-only auditing, while project adaptation requires 7--27 semantic annotations. Together, these results highlight the practical potential of our approach and its advantages over the baselines. They also show that static analysis evidence and source-based auditing complement each other in detecting and assessing real deadlocks.

\bibliographystyle{ACM-Reference-Format}
\bibliography{ref.bib}


\begin{thebibliography}{52}


\ifx \showCODEN    \undefined \def \showCODEN     #1{\unskip}     \fi
\ifx \showDOI      \undefined \def \showDOI       #1{#1}\fi
\ifx \showISBNx    \undefined \def \showISBNx     #1{\unskip}     \fi
\ifx \showISBNxiii \undefined \def \showISBNxiii  #1{\unskip}     \fi
\ifx \showISSN     \undefined \def \showISSN      #1{\unskip}     \fi
\ifx \showLCCN     \undefined \def \showLCCN      #1{\unskip}     \fi
\ifx \shownote     \undefined \def \shownote      #1{#1}          \fi
\ifx \showarticletitle \undefined \def \showarticletitle #1{#1}   \fi
\ifx \showURL      \undefined \def \showURL       {\relax}        \fi
\providecommand\bibfield[2]{#2}
\providecommand\bibinfo[2]{#2}
\providecommand\natexlab[1]{#1}
\providecommand\showeprint[2][]{arXiv:#2}

\bibitem[Astrauskas et~al\mbox{.}(2022)]%
        {astrauskas2022prusti}
\bibfield{author}{\bibinfo{person}{Vytautas Astrauskas}, \bibinfo{person}{Aurel
  B{\'\i}l{\`y}}, \bibinfo{person}{Jon{\'a}{\v{s}} Fiala},
  \bibinfo{person}{Zachary Grannan}, \bibinfo{person}{Christoph Matheja},
  \bibinfo{person}{Peter M{\"u}ller}, \bibinfo{person}{Federico Poli}, {and}
  \bibinfo{person}{Alexander~J Summers}.} \bibinfo{year}{2022}\natexlab{}.
\newblock \showarticletitle{The prusti project: Formal verification for Rust}.
  In \bibinfo{booktitle}{\emph{NASA Formal Methods Symposium}}. Springer,
  \bibinfo{publisher}{Springer-Verlag}, \bibinfo{address}{Berlin, Heidelberg},
  \bibinfo{pages}{88--108}.
\newblock


\bibitem[Astrauskas et~al\mbox{.}(2020)]%
        {astrauskas2020programmers}
\bibfield{author}{\bibinfo{person}{Vytautas Astrauskas},
  \bibinfo{person}{Christoph Matheja}, \bibinfo{person}{Federico Poli},
  \bibinfo{person}{Peter M{\"u}ller}, {and} \bibinfo{person}{Alexander~J
  Summers}.} \bibinfo{year}{2020}\natexlab{}.
\newblock \showarticletitle{How do programmers use unsafe \text{Rust}?}
\newblock \bibinfo{journal}{\emph{Proceedings of the ACM on Programming
  Languages}} \bibinfo{volume}{4}, \bibinfo{number}{OOPSLA}
  (\bibinfo{year}{2020}), \bibinfo{pages}{1--27}.
\newblock


\bibitem[Bae et~al\mbox{.}(2021)]%
        {bae2021rudra}
\bibfield{author}{\bibinfo{person}{Yechan Bae}, \bibinfo{person}{Youngsuk Kim},
  \bibinfo{person}{Ammar Askar}, \bibinfo{person}{Jungwon Lim}, {and}
  \bibinfo{person}{Taesoo Kim}.} \bibinfo{year}{2021}\natexlab{}.
\newblock \showarticletitle{Rudra: Finding memory safety bugs in Rust at the
  ecosystem scale}. In \bibinfo{booktitle}{\emph{Proceedings of the ACM SIGOPS
  28th Symposium on Operating Systems Principles}}.
  \bibinfo{publisher}{Association for Computing Machinery},
  \bibinfo{address}{New York, NY, USA}, \bibinfo{pages}{84--99}.
\newblock


\bibitem[Bai et~al\mbox{.}(2022)]%
        {bai2022dlos}
\bibfield{author}{\bibinfo{person}{Jia-Ju Bai}, \bibinfo{person}{Tuo Li}, {and}
  \bibinfo{person}{Shi-Min Hu}.} \bibinfo{year}{2022}\natexlab{}.
\newblock \showarticletitle{{DLOS}: Effective Static Detection of Deadlocks in
  {OS} Kernels}. In \bibinfo{booktitle}{\emph{2022 USENIX Annual Technical
  Conference (USENIX ATC 22)}}. \bibinfo{publisher}{USENIX Association},
  \bibinfo{address}{Carlsbad, CA}, \bibinfo{pages}{367--382}.
\newblock
\showISBNx{978-1-939133-29-47}
\urldef\tempurl%
\url{https://www.usenix.org/conference/atc22/presentation/bai}
\showURL{%
\tempurl}


\bibitem[Balasubramanian et~al\mbox{.}(2017)]%
        {balasubramanian2017system}
\bibfield{author}{\bibinfo{person}{Abhiram Balasubramanian},
  \bibinfo{person}{Marek~S Baranowski}, \bibinfo{person}{Anton Burtsev},
  \bibinfo{person}{Aurojit Panda}, \bibinfo{person}{Zvonimir Rakamari{\'c}},
  {and} \bibinfo{person}{Leonid Ryzhyk}.} \bibinfo{year}{2017}\natexlab{}.
\newblock \showarticletitle{System programming in rust: Beyond safety}. In
  \bibinfo{booktitle}{\emph{Proceedings of the 16th workshop on hot topics in
  operating systems}}. \bibinfo{pages}{156--161}.
\newblock


\bibitem[Blanc and Lam(2024)]%
        {blanc2024surveying}
\bibfield{author}{\bibinfo{person}{Alex~Le Blanc} {and}
  \bibinfo{person}{Patrick Lam}.} \bibinfo{year}{2024}\natexlab{}.
\newblock \showarticletitle{Surveying the Rust Verification Landscape}.
\newblock \bibinfo{journal}{\emph{arXiv preprint arXiv:2410.01981}}
  (\bibinfo{year}{2024}).
\newblock


\bibitem[Boos et~al\mbox{.}(2020)]%
        {boos2020theseus}
\bibfield{author}{\bibinfo{person}{Kevin Boos}, \bibinfo{person}{Namitha
  Liyanage}, \bibinfo{person}{Ramla Ijaz}, {and} \bibinfo{person}{Lin Zhong}.}
  \bibinfo{year}{2020}\natexlab{}.
\newblock \showarticletitle{Theseus: an experiment in operating system
  structure and state management}. In \bibinfo{booktitle}{\emph{14th USENIX
  Symposium on Operating Systems Design and Implementation (OSDI 20)}}.
  \bibinfo{pages}{1--19}.
\newblock


\bibitem[Corbet(2006)]%
        {corbet2006lockdep}
\bibfield{author}{\bibinfo{person}{Jonathan Corbet}.}
  \bibinfo{year}{2006}\natexlab{}.
\newblock \showarticletitle{The kernel lock validator}.
\newblock \bibinfo{journal}{\emph{Linux Weekly News (LWN)}}
  (\bibinfo{year}{2006}).
\newblock
\urldef\tempurl%
\url{https://lwn.net/Articles/185666/}
\showURL{%
\tempurl}


\bibitem[Cui et~al\mbox{.}(2023)]%
        {cui2023safedrop}
\bibfield{author}{\bibinfo{person}{Mohan Cui}, \bibinfo{person}{Chengjun Chen},
  \bibinfo{person}{Hui Xu}, {and} \bibinfo{person}{Yangfan Zhou}.}
  \bibinfo{year}{2023}\natexlab{}.
\newblock \showarticletitle{SafeDrop: Detecting memory deallocation bugs of
  rust programs via static data-flow analysis}.
\newblock \bibinfo{journal}{\emph{ACM Transactions on Software Engineering and
  Methodology}} \bibinfo{volume}{32}, \bibinfo{number}{4}
  (\bibinfo{year}{2023}), \bibinfo{pages}{1--21}.
\newblock


\bibitem[Cui et~al\mbox{.}(2024)]%
        {cui2024unsafe}
\bibfield{author}{\bibinfo{person}{Mohan Cui}, \bibinfo{person}{Shuran Sun},
  \bibinfo{person}{Hui Xu}, {and} \bibinfo{person}{Yangfan Zhou}.}
  \bibinfo{year}{2024}\natexlab{}.
\newblock \showarticletitle{Is unsafe an Achilles' Heel? A comprehensive study
  of safety requirements in unsafe Rust programming}. In
  \bibinfo{booktitle}{\emph{Proceedings of the IEEE/ACM 46th International
  Conference on Software Engineering}}. \bibinfo{publisher}{Association for
  Computing Machinery}, \bibinfo{address}{New York, NY, USA},
  \bibinfo{pages}{1--13}.
\newblock


\bibitem[Dossche et~al\mbox{.}(2024)]%
        {dossche2024contextsensitiveoutlierbasedstaticanalysis}
\bibfield{author}{\bibinfo{person}{Niels Dossche}, \bibinfo{person}{Bert
  Abrath}, {and} \bibinfo{person}{Bart Coppens}.}
  \bibinfo{year}{2024}\natexlab{}.
\newblock \bibinfo{title}{A Context-Sensitive, Outlier-Based Static Analysis to
  Find Kernel Race Conditions}.
\newblock
\newblock
\showeprint[arxiv]{2404.00350}~[cs.SE]
\urldef\tempurl%
\url{https://arxiv.org/abs/2404.00350}
\showURL{%
\tempurl}


\bibitem[Engler and Ashcraft(2003)]%
        {engler2003racerx}
\bibfield{author}{\bibinfo{person}{Dawson Engler} {and} \bibinfo{person}{Ken
  Ashcraft}.} \bibinfo{year}{2003}\natexlab{}.
\newblock \showarticletitle{RacerX: Effective, Static Detection of Race
  Conditions and Deadlocks}. In \bibinfo{booktitle}{\emph{Proceedings of the
  Nineteenth ACM Symposium on Operating Systems Principles}}.
  \bibinfo{publisher}{Association for Computing Machinery},
  \bibinfo{address}{New York, NY, USA}, \bibinfo{pages}{237--252}.
\newblock
\urldef\tempurl%
\url{https://doi.org/10.1145/945445.945468}
\showDOI{\tempurl}


\bibitem[Evans et~al\mbox{.}(2020)]%
        {evans2020rust}
\bibfield{author}{\bibinfo{person}{Ana~Nora Evans}, \bibinfo{person}{Bradford
  Campbell}, {and} \bibinfo{person}{Mary~Lou Soffa}.}
  \bibinfo{year}{2020}\natexlab{}.
\newblock \showarticletitle{Is \text{Rust} used safely by software
  developers?}. In \bibinfo{booktitle}{\emph{Proceedings of the ACM/IEEE 42nd
  International Conference on Software Engineering}}.
  \bibinfo{publisher}{Association for Computing Machinery},
  \bibinfo{address}{New York, NY, USA}, \bibinfo{pages}{246--257}.
\newblock


\bibitem[Fonseca et~al\mbox{.}(2014)]%
        {fonseca2014ski}
\bibfield{author}{\bibinfo{person}{Pedro Fonseca}, \bibinfo{person}{Rodrigo
  Rodrigues}, {and} \bibinfo{person}{Bj{\"o}rn~B Brandenburg}.}
  \bibinfo{year}{2014}\natexlab{}.
\newblock \showarticletitle{{SKI}: Exposing kernel concurrency bugs through
  systematic schedule exploration}. In \bibinfo{booktitle}{\emph{11th USENIX
  Symposium on Operating Systems Design and Implementation (OSDI)}}.
  \bibinfo{pages}{415--431}.
\newblock


\bibitem[Fu et~al\mbox{.}(2018)]%
        {fu2018survey}
\bibfield{author}{\bibinfo{person}{Haojie Fu}, \bibinfo{person}{Zan Wang},
  \bibinfo{person}{Xiang Chen}, {and} \bibinfo{person}{Xiangyu Fan}.}
  \bibinfo{year}{2018}\natexlab{}.
\newblock \showarticletitle{A systematic survey on automated concurrency bug
  detection, exposing, avoidance, and fixing techniques}.
\newblock \bibinfo{journal}{\emph{Software Quality Journal}}
  \bibinfo{volume}{26} (\bibinfo{year}{2018}), \bibinfo{pages}{855--889}.
\newblock


\bibitem[Fulton et~al\mbox{.}(2021)]%
        {fulton2021benefits}
\bibfield{author}{\bibinfo{person}{Kelsey~R Fulton}, \bibinfo{person}{Anna
  Chan}, \bibinfo{person}{Daniel Votipka}, \bibinfo{person}{Michael Hicks},
  {and} \bibinfo{person}{Michelle~L Mazurek}.} \bibinfo{year}{2021}\natexlab{}.
\newblock \showarticletitle{Benefits and drawbacks of adopting a secure
  programming language: Rust as a case study}. In
  \bibinfo{booktitle}{\emph{Seventeenth Symposium on Usable Privacy and
  Security (SOUPS 2021)}}. \bibinfo{pages}{597--616}.
\newblock


\bibitem[G{\"a}her et~al\mbox{.}(2024)]%
        {gaher2024refinedrust}
\bibfield{author}{\bibinfo{person}{Lennard G{\"a}her}, \bibinfo{person}{Michael
  Sammler}, \bibinfo{person}{Ralf Jung}, \bibinfo{person}{Robbert Krebbers},
  {and} \bibinfo{person}{Derek Dreyer}.} \bibinfo{year}{2024}\natexlab{}.
\newblock \showarticletitle{RefinedRust: A type system for high-assurance
  verification of Rust programs}.
\newblock \bibinfo{journal}{\emph{Proceedings of the ACM on Programming
  Languages}} \bibinfo{volume}{8}, \bibinfo{number}{PLDI}
  (\bibinfo{year}{2024}), \bibinfo{pages}{1115--1139}.
\newblock


\bibitem[Gong et~al\mbox{.}(2021)]%
        {gong2021snowboard}
\bibfield{author}{\bibinfo{person}{Sishuai Gong}, \bibinfo{person}{Deniz
  Altinb{\"u}ken}, \bibinfo{person}{Pedro Fonseca}, {and}
  \bibinfo{person}{Petros Maniatis}.} \bibinfo{year}{2021}\natexlab{}.
\newblock \showarticletitle{Snowboard: Finding kernel concurrency bugs through
  systematic inter-thread communication analysis}. In
  \bibinfo{booktitle}{\emph{Proceedings of the ACM SIGOPS 28th Symposium on
  Operating Systems Principles}}. \bibinfo{pages}{66--83}.
\newblock


\bibitem[Hong and Kim(2013)]%
        {hong2013cobet}
\bibfield{author}{\bibinfo{person}{Shin Hong} {and} \bibinfo{person}{Moonzoo
  Kim}.} \bibinfo{year}{2013}\natexlab{}.
\newblock \showarticletitle{Effective pattern-driven concurrency bug detection
  for operating systems}.
\newblock \bibinfo{journal}{\emph{Journal of Systems and Software}}
  \bibinfo{volume}{86}, \bibinfo{number}{2} (\bibinfo{year}{2013}),
  \bibinfo{pages}{377--388}.
\newblock


\bibitem[Jung et~al\mbox{.}(2017)]%
        {jung2017rustbelt}
\bibfield{author}{\bibinfo{person}{Ralf Jung}, \bibinfo{person}{Jacques-Henri
  Jourdan}, \bibinfo{person}{Robbert Krebbers}, {and} \bibinfo{person}{Derek
  Dreyer}.} \bibinfo{year}{2017}\natexlab{}.
\newblock \showarticletitle{RustBelt: Securing the foundations of the Rust
  programming language}.
\newblock \bibinfo{journal}{\emph{Proceedings of the ACM on Programming
  Languages}} \bibinfo{volume}{2}, \bibinfo{number}{POPL}
  (\bibinfo{year}{2017}), \bibinfo{pages}{1--34}.
\newblock


\bibitem[Jung et~al\mbox{.}(2021)]%
        {jung2021safe}
\bibfield{author}{\bibinfo{person}{Ralf Jung}, \bibinfo{person}{Jacques-Henri
  Jourdan}, \bibinfo{person}{Robbert Krebbers}, {and} \bibinfo{person}{Derek
  Dreyer}.} \bibinfo{year}{2021}\natexlab{}.
\newblock \showarticletitle{Safe Systems Programming in Rust}.
\newblock \bibinfo{journal}{\emph{Commun. ACM}} \bibinfo{volume}{64},
  \bibinfo{number}{4} (\bibinfo{year}{2021}), \bibinfo{pages}{144--152}.
\newblock
\urldef\tempurl%
\url{https://doi.org/10.1145/3418295}
\showDOI{\tempurl}


\bibitem[Lattuada et~al\mbox{.}(2024)]%
        {lattuada2024verus}
\bibfield{author}{\bibinfo{person}{Andrea Lattuada}, \bibinfo{person}{Travis
  Hance}, \bibinfo{person}{Jay Bosamiya}, \bibinfo{person}{Matthias Brun},
  \bibinfo{person}{Chanhee Cho}, \bibinfo{person}{Hayley LeBlanc},
  \bibinfo{person}{Pranav Srinivasan}, \bibinfo{person}{Reto Achermann},
  \bibinfo{person}{Tej Chajed}, \bibinfo{person}{Chris Hawblitzel},
  {et~al\mbox{.}}} \bibinfo{year}{2024}\natexlab{}.
\newblock \showarticletitle{Verus: A practical foundation for systems
  verification}. In \bibinfo{booktitle}{\emph{Proceedings of the ACM SIGOPS
  30th Symposium on Operating Systems Principles}}.
  \bibinfo{publisher}{Association for Computing Machinery},
  \bibinfo{address}{New York, NY, USA}, \bibinfo{pages}{438--454}.
\newblock


\bibitem[Lattuada et~al\mbox{.}(2023)]%
        {lattuada2023verus}
\bibfield{author}{\bibinfo{person}{Andrea Lattuada}, \bibinfo{person}{Travis
  Hance}, \bibinfo{person}{Chanhee Cho}, \bibinfo{person}{Matthias Brun},
  \bibinfo{person}{Isitha Subasinghe}, \bibinfo{person}{Yi Zhou},
  \bibinfo{person}{Jon Howell}, \bibinfo{person}{Bryan Parno}, {and}
  \bibinfo{person}{Chris Hawblitzel}.} \bibinfo{year}{2023}\natexlab{}.
\newblock \showarticletitle{Verus: Verifying Rust programs using linear ghost
  types}.
\newblock \bibinfo{journal}{\emph{Proceedings of the ACM on Programming
  Languages}} \bibinfo{volume}{7}, \bibinfo{number}{OOPSLA1}
  (\bibinfo{year}{2023}), \bibinfo{pages}{286--315}.
\newblock


\bibitem[Lehmann et~al\mbox{.}(2023)]%
        {lehmann2023flux}
\bibfield{author}{\bibinfo{person}{Nico Lehmann}, \bibinfo{person}{Adam~T
  Geller}, \bibinfo{person}{Niki Vazou}, {and} \bibinfo{person}{Ranjit Jhala}.}
  \bibinfo{year}{2023}\natexlab{}.
\newblock \showarticletitle{Flux: Liquid types for rust}.
\newblock \bibinfo{journal}{\emph{Proceedings of the ACM on Programming
  Languages}} \bibinfo{volume}{7}, \bibinfo{number}{PLDI}
  (\bibinfo{year}{2023}), \bibinfo{pages}{1533--1557}.
\newblock


\bibitem[Li et~al\mbox{.}(2023)]%
        {li2023embedded}
\bibfield{author}{\bibinfo{person}{Chao Li}, \bibinfo{person}{Rui Chen},
  \bibinfo{person}{Boxiang Wang}, \bibinfo{person}{Zhixuan Wang},
  \bibinfo{person}{Tingting Yu}, \bibinfo{person}{Yunsong Jiang},
  \bibinfo{person}{Bin Gu}, {and} \bibinfo{person}{Mengfei Yang}.}
  \bibinfo{year}{2023}\natexlab{}.
\newblock \showarticletitle{An empirical study on concurrency bugs in
  interrupt-driven embedded software}. In \bibinfo{booktitle}{\emph{Proceedings
  of the 32nd ACM SIGSOFT International Symposium on Software Testing and
  Analysis}}. \bibinfo{pages}{1345--1356}.
\newblock


\bibitem[Li et~al\mbox{.}(2022)]%
        {li2022atomicity}
\bibfield{author}{\bibinfo{person}{Chao Li}, \bibinfo{person}{Rui Chen},
  \bibinfo{person}{Boxiang Wang}, \bibinfo{person}{Tingting Yu},
  \bibinfo{person}{Dongdong Gao}, {and} \bibinfo{person}{Mengfei Yang}.}
  \bibinfo{year}{2022}\natexlab{}.
\newblock \showarticletitle{Precise and efficient atomicity violation detection
  for interrupt-driven programs via staged path pruning}. In
  \bibinfo{booktitle}{\emph{Proceedings of the 31st ACM SIGSOFT International
  Symposium on Software Testing and Analysis}}. \bibinfo{pages}{506--518}.
\newblock


\bibitem[Li et~al\mbox{.}(2024)]%
        {li2024empirical}
\bibfield{author}{\bibinfo{person}{Hongyu Li}, \bibinfo{person}{Liwei Guo},
  \bibinfo{person}{Yexuan Yang}, \bibinfo{person}{Shangguang Wang}, {and}
  \bibinfo{person}{Mengwei Xu}.} \bibinfo{year}{2024}\natexlab{}.
\newblock \showarticletitle{An empirical study of Rust-for-Linux: The success,
  dissatisfaction, and compromise}. In \bibinfo{booktitle}{\emph{2024 USENIX
  Annual Technical Conference}}. \bibinfo{publisher}{USENIX Association},
  \bibinfo{address}{Santa Clara, CA}, \bibinfo{pages}{425--443}.
\newblock


\bibitem[Li et~al\mbox{.}(2021)]%
        {li2021mirchecker}
\bibfield{author}{\bibinfo{person}{Zhuohua Li}, \bibinfo{person}{Jincheng
  Wang}, \bibinfo{person}{Mingshen Sun}, {and} \bibinfo{person}{John~CS Lui}.}
  \bibinfo{year}{2021}\natexlab{}.
\newblock \showarticletitle{MirChecker: detecting bugs in Rust programs via
  static analysis}. In \bibinfo{booktitle}{\emph{Proceedings of The 2021 ACM
  SIGSAC Conference on Computer and Communications Security}}.
  \bibinfo{publisher}{Association for Computing Machinery},
  \bibinfo{address}{New York, NY, USA}, \bibinfo{pages}{2183--2196}.
\newblock


\bibitem[Liu et~al\mbox{.}(2020)]%
        {liu2020securing}
\bibfield{author}{\bibinfo{person}{Peiming Liu}, \bibinfo{person}{Gang Zhao},
  {and} \bibinfo{person}{Jeff Huang}.} \bibinfo{year}{2020}\natexlab{}.
\newblock \showarticletitle{Securing unsafe rust programs with XRust}. In
  \bibinfo{booktitle}{\emph{Proceedings of the ACM/IEEE 42nd International
  Conference on Software Engineering}}. \bibinfo{pages}{234--245}.
\newblock


\bibitem[Lu et~al\mbox{.}(2008)]%
        {lu2008learning}
\bibfield{author}{\bibinfo{person}{Shan Lu}, \bibinfo{person}{Soyeon Park},
  \bibinfo{person}{Eunsoo Seo}, {and} \bibinfo{person}{Yuanyuan Zhou}.}
  \bibinfo{year}{2008}\natexlab{}.
\newblock \showarticletitle{Learning from mistakes: a comprehensive study on
  real world concurrency bug characteristics}. In
  \bibinfo{booktitle}{\emph{Proceedings of the 13th International Conference on
  Architectural Support for Programming Languages and Operating Systems}}.
  \bibinfo{pages}{329--339}.
\newblock


\bibitem[Matsushita et~al\mbox{.}(2022)]%
        {matsushita2022rusthornbelt}
\bibfield{author}{\bibinfo{person}{Yusuke Matsushita}, \bibinfo{person}{Xavier
  Denis}, \bibinfo{person}{Jacques-Henri Jourdan}, {and} \bibinfo{person}{Derek
  Dreyer}.} \bibinfo{year}{2022}\natexlab{}.
\newblock \showarticletitle{RustHornBelt: a semantic foundation for functional
  verification of Rust programs with unsafe code}. In
  \bibinfo{booktitle}{\emph{Proceedings of the 43rd ACM SIGPLAN International
  Conference on Programming Language Design and Implementation}}.
  \bibinfo{publisher}{Association for Computing Machinery},
  \bibinfo{address}{New York, NY, USA}, \bibinfo{pages}{841--856}.
\newblock


\bibitem[Narayanan et~al\mbox{.}(2020)]%
        {narayanan2020redleaf}
\bibfield{author}{\bibinfo{person}{Vikram Narayanan}, \bibinfo{person}{Tianjiao
  Huang}, \bibinfo{person}{David Detweiler}, \bibinfo{person}{Dan Appel},
  \bibinfo{person}{Zhaofeng Li}, \bibinfo{person}{Gerd Zellweger}, {and}
  \bibinfo{person}{Anton Burtsev}.} \bibinfo{year}{2020}\natexlab{}.
\newblock \showarticletitle{RedLeaf: isolation and communication in a safe
  operating system}. In \bibinfo{booktitle}{\emph{14th USENIX Symposium on
  Operating Systems Design and Implementation (OSDI 20)}}.
  \bibinfo{pages}{21--39}.
\newblock


\bibitem[Ning and Qin(2020)]%
        {ning2020stuckmenot}
\bibfield{author}{\bibinfo{person}{Pengxiang Ning} {and} \bibinfo{person}{Boqin
  Qin}.} \bibinfo{year}{2020}\natexlab{}.
\newblock \showarticletitle{Stuck-me-not: A Deadlock Detector on Blockchain
  Software in Rust}.
\newblock \bibinfo{journal}{\emph{Procedia Computer Science}}
  \bibinfo{volume}{177} (\bibinfo{year}{2020}), \bibinfo{pages}{599--604}.
\newblock
\urldef\tempurl%
\url{https://doi.org/10.1016/j.procs.2020.10.085}
\showDOI{\tempurl}


\bibitem[Peng et~al\mbox{.}(2024)]%
        {peng2024framekernel}
\bibfield{author}{\bibinfo{person}{Yuke Peng}, \bibinfo{person}{Hongliang
  Tian}, \bibinfo{person}{Jinyi Xian}, \bibinfo{person}{Shuai Zhou},
  \bibinfo{person}{Shoumeng Yan}, {and} \bibinfo{person}{Yinqian Zhang}.}
  \bibinfo{year}{2024}\natexlab{}.
\newblock \showarticletitle{Framekernel: A safe and efficient kernel
  architecture via Rust-based intra-kernel privilege separation}. In
  \bibinfo{booktitle}{\emph{Proceedings of the 15th ACM SIGOPS Asia-Pacific
  Workshop on Systems}}. \bibinfo{publisher}{Association for Computing
  Machinery}, \bibinfo{address}{New York, NY, USA}, \bibinfo{pages}{31--37}.
\newblock


\bibitem[Peng et~al\mbox{.}(2025)]%
        {peng2025asterinas}
\bibfield{author}{\bibinfo{person}{Yuke Peng}, \bibinfo{person}{Hongliang
  Tian}, \bibinfo{person}{Junyang Zhang}, \bibinfo{person}{Ruihan Li},
  \bibinfo{person}{Chengjun Chen}, \bibinfo{person}{Jianfeng Jiang},
  \bibinfo{person}{Jinyi Xian}, \bibinfo{person}{Xiaolin Wang},
  \bibinfo{person}{Chenren Xu}, \bibinfo{person}{Diyu Zhou},
  \bibinfo{person}{Yingwei Luo}, \bibinfo{person}{Shoumeng Yan}, {and}
  \bibinfo{person}{Yinqian Zhang}.} \bibinfo{year}{2025}\natexlab{}.
\newblock \showarticletitle{ASTERINAS: a Linux ABI-compatible, rust-based
  framekernel OS with a small and sound TCB}. In
  \bibinfo{booktitle}{\emph{Proceedings of the 2025 USENIX Conference on Usenix
  Annual Technical Conference}} (Boston, MA, USA)
  \emph{(\bibinfo{series}{USENIX ATC '25})}. \bibinfo{publisher}{USENIX
  Association}, \bibinfo{address}{USA}, Article \bibinfo{articleno}{19},
  \bibinfo{numpages}{17}~pages.
\newblock
\showISBNx{978-1-939133-48-9}


\bibitem[Qin et~al\mbox{.}(2024)]%
        {qin2024understanding}
\bibfield{author}{\bibinfo{person}{Boqin Qin}, \bibinfo{person}{Yilun Chen},
  \bibinfo{person}{Haopeng Liu}, \bibinfo{person}{Hua Zhang},
  \bibinfo{person}{Qiaoyan Wen}, \bibinfo{person}{Linhai Song}, {and}
  \bibinfo{person}{Yiying Zhang}.} \bibinfo{year}{2024}\natexlab{}.
\newblock \showarticletitle{Understanding and detecting real-world safety
  issues in Rust}.
\newblock \bibinfo{journal}{\emph{IEEE Transactions on Software Engineering}}
  \bibinfo{volume}{50}, \bibinfo{number}{6} (\bibinfo{year}{2024}),
  \bibinfo{pages}{1306--1324}.
\newblock


\bibitem[Qin et~al\mbox{.}(2020)]%
        {qin2020rust}
\bibfield{author}{\bibinfo{person}{Boqin Qin}, \bibinfo{person}{Yilun Chen},
  \bibinfo{person}{Zeming Yu}, \bibinfo{person}{Linhai Song}, {and}
  \bibinfo{person}{Yiying Zhang}.} \bibinfo{year}{2020}\natexlab{}.
\newblock \showarticletitle{Understanding memory and thread safety practices
  and issues in real-world Rust programs}. In
  \bibinfo{booktitle}{\emph{Proceedings of the 41st ACM SIGPLAN Conference on
  Programming Language Design and Implementation}}. \bibinfo{pages}{763--779}.
\newblock


\bibitem[Rao(2024)]%
        {Rao2024SemiAutomatedVO}
\bibfield{author}{\bibinfo{person}{Zihao Rao}.}
  \bibinfo{year}{2024}\natexlab{}.
\newblock \showarticletitle{Semi-Automated Verification of Interior Unsafe Code
  Encapsulation in Real-World Rust Systems}.
\newblock \bibinfo{journal}{\emph{2024 39th IEEE/ACM International Conference
  on Automated Software Engineering (ASE)}} (\bibinfo{year}{2024}),
  \bibinfo{pages}{2435--2437}.
\newblock
\urldef\tempurl%
\url{https://api.semanticscholar.org/CorpusID:273478731}
\showURL{%
\tempurl}


\bibitem[Rao et~al\mbox{.}(2025)]%
        {rao2025annotating}
\bibfield{author}{\bibinfo{person}{Zihao Rao}, \bibinfo{person}{Hongliang
  Tian}, \bibinfo{person}{Xin Wang}, {and} \bibinfo{person}{Hui Xu}.}
  \bibinfo{year}{2025}\natexlab{}.
\newblock \showarticletitle{Annotating and Auditing the Safety Properties of
  Unsafe Rust}.
\newblock \bibinfo{journal}{\emph{arXiv preprint arXiv:2504.21312}}
  (\bibinfo{year}{2025}).
\newblock


\bibitem[Rao et~al\mbox{.}(2024)]%
        {rao2024characterizing}
\bibfield{author}{\bibinfo{person}{Zihao Rao}, \bibinfo{person}{Yiran Yang},
  {and} \bibinfo{person}{Hui Xu}.} \bibinfo{year}{2024}\natexlab{}.
\newblock \showarticletitle{Characterizing Unsafe Code Encapsulation In
  Real-world Rust Systems}.
\newblock \bibinfo{journal}{\emph{arXiv preprint arXiv:2406.07936}}
  (\bibinfo{year}{2024}).
\newblock


\bibitem[Rivera et~al\mbox{.}(2021)]%
        {rivera2021keeping}
\bibfield{author}{\bibinfo{person}{Elijah Rivera}, \bibinfo{person}{Samuel
  Mergendahl}, \bibinfo{person}{Howard Shrobe}, \bibinfo{person}{Hamed
  Okhravi}, {and} \bibinfo{person}{Nathan Burow}.}
  \bibinfo{year}{2021}\natexlab{}.
\newblock \showarticletitle{Keeping safe rust safe with galeed}. In
  \bibinfo{booktitle}{\emph{Proceedings of the 37th Annual Computer Security
  Applications Conference}}. \bibinfo{pages}{824--836}.
\newblock


\bibitem[Russell(nd)]%
        {linuxlockdoc}
\bibfield{author}{\bibinfo{person}{Rusty Russell}.}
  \bibinfo{year}{{n.d.}}\natexlab{}.
\newblock \bibinfo{title}{Unreliable Guide To Locking}.
\newblock
\newblock
\urldef\tempurl%
\url{https://docs.kernel.org/6.6/kernel-hacking/locking.html}
\showURL{%
\tempurl}


\bibitem[Santhiar and Kanade(2017)]%
        {santhiar2017deadwait}
\bibfield{author}{\bibinfo{person}{Anirudh Santhiar} {and}
  \bibinfo{person}{Aditya Kanade}.} \bibinfo{year}{2017}\natexlab{}.
\newblock \showarticletitle{Static deadlock detection for asynchronous C\#
  programs}.
\newblock \bibinfo{journal}{\emph{SIGPLAN Not.}} \bibinfo{volume}{52},
  \bibinfo{number}{6} (\bibinfo{date}{June} \bibinfo{year}{2017}),
  \bibinfo{pages}{292–305}.
\newblock
\showISSN{0362-1340}
\urldef\tempurl%
\url{https://doi.org/10.1145/3140587.3062361}
\showDOI{\tempurl}


\bibitem[Savage et~al\mbox{.}(1997)]%
        {savage1997eraser}
\bibfield{author}{\bibinfo{person}{Stefan Savage}, \bibinfo{person}{Michael
  Burrows}, \bibinfo{person}{Greg Nelson}, \bibinfo{person}{Patrick
  Sobalvarro}, {and} \bibinfo{person}{Thomas Anderson}.}
  \bibinfo{year}{1997}\natexlab{}.
\newblock \showarticletitle{Eraser: a dynamic data race detector for
  multithreaded programs}.
\newblock \bibinfo{journal}{\emph{ACM Trans. Comput. Syst.}}
  \bibinfo{volume}{15}, \bibinfo{number}{4} (\bibinfo{date}{Nov.}
  \bibinfo{year}{1997}), \bibinfo{pages}{391–411}.
\newblock
\showISSN{0734-2071}
\urldef\tempurl%
\url{https://doi.org/10.1145/265924.265927}
\showDOI{\tempurl}


\bibitem[Serebryany and Iskhodzhanov(2009)]%
        {serebryany2009tsan}
\bibfield{author}{\bibinfo{person}{Konstantin Serebryany} {and}
  \bibinfo{person}{Timur Iskhodzhanov}.} \bibinfo{year}{2009}\natexlab{}.
\newblock \showarticletitle{ThreadSanitizer: data race detection in practice}.
  In \bibinfo{booktitle}{\emph{Proceedings of the Workshop on Binary
  Instrumentation and Applications}} (New York, New York, USA)
  \emph{(\bibinfo{series}{WBIA '09})}. \bibinfo{publisher}{Association for
  Computing Machinery}, \bibinfo{address}{New York, NY, USA},
  \bibinfo{pages}{62–71}.
\newblock
\showISBNx{9781605587936}
\urldef\tempurl%
\url{https://doi.org/10.1145/1791194.1791203}
\showDOI{\tempurl}


\bibitem[Sharma et~al\mbox{.}(2024)]%
        {sharma2024rust}
\bibfield{author}{\bibinfo{person}{Ayushi Sharma}, \bibinfo{person}{Shashank
  Sharma}, \bibinfo{person}{Sai~Ritvik Tanksalkar}, \bibinfo{person}{Santiago
  Torres-Arias}, {and} \bibinfo{person}{Aravind Machiry}.}
  \bibinfo{year}{2024}\natexlab{}.
\newblock \showarticletitle{Rust for Embedded Systems: Current State and Open
  Problems}. In \bibinfo{booktitle}{\emph{Proceedings of the 2024 on ACM SIGSAC
  Conference on Computer and Communications Security}}.
  \bibinfo{publisher}{Association for Computing Machinery},
  \bibinfo{address}{New York, NY, USA}, \bibinfo{pages}{2296--2310}.
\newblock


\bibitem[{The Rust Project Developers}(2026)]%
        {rustReferenceDestructors}
\bibfield{author}{\bibinfo{person}{{The Rust Project Developers}}.}
  \bibinfo{year}{2026}\natexlab{}.
\newblock \bibinfo{title}{The Rust Reference: Destructors and Drop Scopes}.
\newblock
\newblock
\urldef\tempurl%
\url{https://doc.rust-lang.org/reference/destructors.html}
\showURL{%
\tempurl}


\bibitem[Tian et~al\mbox{.}(2025)]%
        {tian2025asterinas}
\bibfield{author}{\bibinfo{person}{Hongliang Tian}, \bibinfo{person}{Yuke
  Peng}, \bibinfo{person}{Yingwei Luo}, \bibinfo{person}{Shoumeng Yan}, {and}
  \bibinfo{person}{Yinqian Zhang}.} \bibinfo{year}{2025}\natexlab{}.
\newblock \showarticletitle{Asterinas: A Rust-Based Framekernel to Reimagine
  Linux in the 2020s}.
\newblock  (\bibinfo{year}{2025}).
\newblock
\urldef\tempurl%
\url{https://www.usenix.org/publications/loginonline/asterinas-rust-based-framekernel-reimagine-linux-2020s}
\showURL{%
\tempurl}


\bibitem[Voung et~al\mbox{.}(2007)]%
        {voung2007relay}
\bibfield{author}{\bibinfo{person}{Jan~Wen Voung}, \bibinfo{person}{Ranjit
  Jhala}, {and} \bibinfo{person}{Sorin Lerner}.}
  \bibinfo{year}{2007}\natexlab{}.
\newblock \showarticletitle{Relay: static race detection on millions of lines
  of code}. In \bibinfo{booktitle}{\emph{Proceedings of the 6th Joint Meeting
  of the European Software Engineering Conference and the ACM SIGSOFT Symposium
  on the Foundations of Software Engineering}}. \bibinfo{pages}{205--214}.
\newblock


\bibitem[Xu et~al\mbox{.}(2021)]%
        {xu2021memory}
\bibfield{author}{\bibinfo{person}{Hui Xu}, \bibinfo{person}{Zhuangbin Chen},
  \bibinfo{person}{Mingshen Sun}, \bibinfo{person}{Yangfan Zhou}, {and}
  \bibinfo{person}{Michael~R Lyu}.} \bibinfo{year}{2021}\natexlab{}.
\newblock \showarticletitle{Memory-safety challenge considered solved? An
  in-depth study with all Rust CVEs}.
\newblock \bibinfo{journal}{\emph{ACM Transactions on Software Engineering and
  Methodology (TOSEM)}} \bibinfo{volume}{31}, \bibinfo{number}{1}
  (\bibinfo{year}{2021}), \bibinfo{pages}{1--25}.
\newblock


\bibitem[Ye et~al\mbox{.}(2024)]%
        {ye2024archerfish}
\bibfield{author}{\bibinfo{person}{Chengfeng Ye}, \bibinfo{person}{Yuandao
  Cai}, {and} \bibinfo{person}{Charles Zhang}.}
  \bibinfo{year}{2024}\natexlab{}.
\newblock \showarticletitle{When Threads Meet Interrupts: Effective Static
  Detection of {Interrupt-Based} Deadlocks in Linux}. In
  \bibinfo{booktitle}{\emph{33rd USENIX Security Symposium (USENIX Security
  24)}}. \bibinfo{publisher}{USENIX Association},
  \bibinfo{address}{Philadelphia, PA}, \bibinfo{pages}{6167--6184}.
\newblock
\showISBNx{978-1-939133-44-1}
\urldef\tempurl%
\url{https://www.usenix.org/conference/usenixsecurity24/presentation/ye}
\showURL{%
\tempurl}


\bibitem[Zhang et~al\mbox{.}(2024)]%
        {zhang2024rustdeadlock}
\bibfield{author}{\bibinfo{person}{Yu Zhang}, \bibinfo{person}{Kaiwen Zhang},
  {and} \bibinfo{person}{Guanjun Liu}.} \bibinfo{year}{2024}\natexlab{}.
\newblock \showarticletitle{Static Deadlock Detection for Rust Programs}.
\newblock \bibinfo{journal}{\emph{CoRR}}  \bibinfo{volume}{abs/2401.01114}
  (\bibinfo{year}{2024}).
\newblock
\urldef\tempurl%
\url{https://doi.org/10.48550/arXiv.2401.01114}
\showDOI{\tempurl}


\end{thebibliography}

\end{document}